\documentclass[a4paper,11pt]{article}
\pdfoutput=0 

\usepackage{jheppub} 

\usepackage[T1]{fontenc} 
\usepackage[normalem]{ulem}
\usepackage{multirow}
\usepackage{bbold}
\def\Id{\mathbb{1}}

\begin{document}

\title{Chiral condensate in the Schwinger model with Matrix Product Operators}

 \author[a]{Mari Carmen Ba\~nuls,}
 \emailAdd{banulsm@mpq.mpg.de} 
 \affiliation[a]{Max-Planck-Institut f\"ur Quantenoptik,
 Hans-Kopfermann-Str. 1, 85748 Garching, Germany}
 \author[b,c]{Krzysztof Cichy,} 
 \affiliation[b]{Goethe-Universit\"at Frankfurt am Main, Institut f\"ur Theoretische Physik,
 Max-von-Laue-Stra\ss e 1, 60438 Frankfurt am Main, Germany}
 \affiliation[c]{Adam Mickiewicz University, Faculty
of Physics,
Umultowska 85, 61-614 Poznan, Poland}
 \emailAdd{kcichy@th.physik.uni-frankfurt.de}
\author[d]{Karl Jansen} 
\affiliation[d]{NIC, DESY Zeuthen, Platanenallee 6, 15738 Zeuthen, Germany}
\emailAdd{Karl.Jansen@desy.de}
\author[e]{and Hana Saito}
\affiliation[e]{Center for Computational Sciences, University of Tsukuba, Ibaraki 305-8577, Japan}
\emailAdd{saitouh@ccs.tsukuba.ac.jp}

\abstract{
Tensor network (TN) methods, in particular the Matrix Product States (MPS) ansatz, have proven to be a useful tool in analyzing the properties of lattice gauge theories.
They allow for a very good precision, much better than standard Monte Carlo (MC) techniques for the models that have been studied so far, due to the possibility of reaching much smaller lattice spacings.
The real reason for the interest in the TN approach, however, is its ability, shown so far in several condensed matter models, to deal with theories which exhibit the notorious sign problem in MC simulations.
This makes it prospective for dealing with the non-zero chemical potential in QCD and other lattice gauge theories, as well as with real-time simulations.
In this paper, using matrix product operators, we extend our analysis of the Schwinger model at zero temperature to show the feasibility of this approach also at finite temperature.
This is an important step on the way to deal with the sign problem of QCD.
We analyze in detail the chiral symmetry breaking in the massless and massive cases and show that the method works very well and gives good control over a broad range of temperatures, essentially from zero to infinite temperature.
}

\preprint{DESY 16-046}


\maketitle
\flushbottom

\section{Introduction}
\label{sec:intro}
Investigations of gauge field theories within the Hamiltonian approach have progressed substantially in the 
last years with the help of tensor network (TN) techniques~\cite{verstraete08algo,cirac09rg,orus2014review}.
Taking the example of the Schwinger model, numerical calculations have been performed to investigate 
ground state properties \cite{Byrnes:2002nv,Cichy:2012rw,Banuls:2013jaa,Banuls:2013zva,Rico:2013qya,Buyens:2015dkc},
to demonstrate real-time dynamics \cite{Buyens:2013yza,Buyens:2014pga}
and to address the phenomenon of string breaking  \cite{Pichler:2015yqa,Buyens:2015tea},
which has also been explored in non-Abelian models \cite{Kuhn:2015zqa}.
In Refs.~\cite{Banuls:2015sta,Saito:2014bda,Saito:2015ryj}, thermal properties of the 
Schwinger model were studied for massless fermions.
From a more conceptual point of view, TN have been developed that incorporate the gauge 
symmetry by construction, and constitute ground states of gauge invariant lattice 
models \cite{Tagliacozzo:2014bta,Silvi:2014pta,haegeman15gauging,zohar2015peps}.
Yet a different line of work is the study of potential quantum simulations of these models,
using ultracold atoms, see Refs.~\cite{wiese2013review,Zohar:2015hwa,Dalmonte:2016alw} for a review.
Also in this field, TN techniques can play a determinant role to study the feasibility of the proposals \cite{kuehn2014schwinger}.

The last numerical developments go beyond standard Markov Chain Monte Carlo (MC-MC) methods. 
At zero temperature, the Hamiltonian approach allows us to go substantially closer to the continuum limit and reach a much improved accuracy compared to MC-MC. 
When temperature is switched on, a broad and very large set of non-zero temperature points can be evaluated, ranging from very high to almost zero temperature.
In the string breaking calculation, a nice picture of the string breaking phenomenon and the emergence of the hadron states can be demonstrated. 
Finally, real-time simulations are not even possible in principle with MC-MC methods. 

The key to this success is the employment of tensor network states and, in the case of one spatial dimension, as for the Schwinger model, the Matrix 
Product States (MPS). 
In this approach, which is closely linked to the Density Matrix Renormalization Group (DMRG) \cite{white92dmrg}, 
the problem, which has an exponentially large dimension in terms of the system size, is reduced to an --admittedly-- sophisticated variational solution which can be
encoded in substantially smaller $D\times D$ matrices.
The ansatz can represent arbitrary states in the Hilbert space if $D$ is large enough (exponential in the system size).
Instead in numerical applications, usually an approximation is found to the desired state within the set of MPS with fixed $D$.
By varying $D$, an extrapolation of results to $D\to \infty$ can be performed allowing thus to reach the solution 
of the real system under consideration. 
A different approach also using tensor network techniques was applied to the Schwinger model with a topological
$\theta$-term in Refs.~\cite{Shimizu:2014uva,Shimizu:2014fsa}, where the exact partition function on the lattice was 
expressed as a two dimensional tensor network and approximately contracted using the Tensor Renormalization Group (TRG).

The application of the MPS technique discussed in the present paper is concerned with non-zero temperature 
properties of the Schwinger model. 
In Refs.~\cite{Banuls:2015sta,Saito:2014bda,Saito:2015ryj}, we have for the first time investigated the thermal evolution of the chiral condensate 
in the Schwinger model. In the first paper, where we only studied the massless case, we could demonstrate that the MPS technique 
can be successfully used to compute such a thermal evolution from very high to almost zero temperature. For massless fermions, the 
results from our MPS calculation could be confronted with the analytical solution of Ref.~\cite{Sachs:1991en} and a very nice
agreement was found demonstrating the correctness and the power of the MPS approach. 

In the present paper, we will extend our calculations of the thermal evolution of the chiral condensate to the case of non-vanishing fermion masses. 
Here, no exact results exist anymore, but only approximate solutions are available \cite{Hosotani:1998za} which can be tested against our results. 
For our work at zero fermion mass, we also introduced a truncation of the charge sector \cite{Banuls:2015sta} which was necessary 
to obtain precise results at high temperature. Here, we will employ this truncation method, too. 

It needs to be stressed that the calculations with MPS, as performed here, have a number of systematic uncertainties which are very important to control. 
This concerns in particular:
\begin{itemize}
\item an estimate of results for infinite bond dimension; \footnote{For a given system size, $N$, exact results would actually be attained with finite bond dimension, $D=2^{N/2}$\cite{verstraete04dmrg}, which is many orders of magnitude larger than the largest one we use in the simulations.}
\item an extrapolation to zero step size in the thermal evolution process;
\item a study of the truncation in the charge sector of the model;
\item an infinite volume extrapolation; 
\item and a careful analysis of the continuum limit employing 
various extrapolation functions with different orders in the lattice spacing.
\end{itemize}

Controlling these systematic effects renders the calculations with MPS demanding, but it is absolutely necessary to obtain precise and trustworthy results. We have therefore made a significant effort to perform the above extrapolations and we will provide various examples in this paper for the 
studies of systematic effects carried through here.

\section{The Schwinger model and chiral symmetry breaking}
\label{sec:schwinger}
The one-flavour Schwinger model \cite{schwinger62}, i.e. Quantum Electrodynamics in 1+1 dimensions, is one of the simplest gauge theories and a toy model allowing for studies of new lattice techniques before employing them to real theories of interest, like Quantum Chromodynamics (QCD).
Despite its apparent simplicity, it has a non-perturbatively generated mass gap and shares some features with QCD, such as confinement and chiral symmetry breaking, although the mechanism of the latter is different than in QCD -- it is not spontaneous, but results from the chiral anomaly.

We start with the Hamiltonian of the Schwinger model in the staggered discretization, derived and discussed in Ref.~\cite{Banks:1975gq}: 
\begin{eqnarray}
  \label{eq:H}
  H
  &=& x \displaystyle \sum_{n=0}^{N-2}
  \left[ \sigma_n^+ \sigma_{n+1}^- 
  + \sigma_n^- \sigma_{n+1}^+ \right]
  +\frac{\mu}{2} \sum_{n=0}^{N-1} 
  \Big[ 1+ (-1)^n \sigma_n^z \Big]
  + \sum_{n=0}^{N-2} \left[ L(n) \right] ^2\\
  &\equiv& H_{hop} + H_m + H_g,\nonumber
  \end{eqnarray}
where $n$ is the site index, $x=1/g^2a^2$, $a$ is the lattice spacing, $g$ is the coupling, and $\mu=2m/g^2a$ with $m$ denoting the fermion mass and $N$ the number of lattice sites.
We use open boundary conditions (OBC).
The gauge field, $L(n)$, can be integrated out using the Gauss law: 
\begin{equation}
   L(n+1) - L(n) = \frac{1}{2} \left[ (-1)^{n+1} + \sigma_{n+1}^z \right].
   \label{eq:Gausslaw}
\end{equation}
Thus, only $L(n)$ at one of the boundaries is an independent parameter and we take $L(0)=0$, i.e. no background electric field.

We work with the following basis for our numerical computations:
$\left| s_0 s_1 \cdots \right\rangle$ ~\cite{Banuls:2013jaa},
where $s_n=\{\downarrow,\uparrow\}$ is the spin state at site $n$ and all the gauge degrees of freedom have 
been integrated out.

In this paper, we are interested in the chiral symmetry breaking ($\chi$SB) in the Schwinger model, both at zero and non-zero temperature.
The order parameter of $\chi$SB is the chiral condensate $\Sigma=\left\langle {\bar \psi}\psi \right\rangle$, 
which can be written in terms of spin operators as $\Sigma = \frac{g\sqrt{x}}{N} \sum_n (-1)^n \frac{1+\sigma_n^z}{2}$.
The ground state and thermal expectation values of the chiral condensate
diverge logarithmically in the continuum limit for non-zero fermion mass~\cite{deForcrand98,duerr05scaling,Christian:2005yp}.
This divergence is present even in the non-interacting case, where the theory is exactly solvable and the Hamiltonian (\ref{eq:H}) reduces to the XY spin model in a staggered magnetic field.
The ground state energy of this model (with OBC) reads:
$\frac{E_0}{g}=\frac{\mu}{2}N-\sum_{q=1}^{N/2} \sqrt{\mu^2+4 x^2 \cos^2\frac{q \pi}{N+1}}$.
The ground state expectation value of $\Sigma$ can then be computed from the derivative $\frac{d E_0}{d\mu}$:
\begin{equation}
\Sigma_{\mathrm{free}}(\mu,x,N)=\frac{g\sqrt{x}}{N}\sum_{q=1}^{N/2}\frac{\mu}{\sqrt{\mu^2+4 x^2 \cos^2 \frac{q \pi}{N+1}}}.
\label{eq:freecond}
\end{equation}
The free condensate value computed from this formula can be used to subtract the divergence in the interacting case at a finite lattice size $N$, a finite lattice spacing $1/\sqrt{x}$ and a given fermion mass $m/g$.
However, one can exactly evaluate the infinite volume limit of the free condensate first, yielding:
\begin{equation}
\Sigma_{\mathrm{free}}(m/g,x)=\frac{m}{\pi}\frac{1}{\sqrt{1+\frac{m^2}{g^2 x}}} \mathrm{K}\left(\frac{1}{1+\frac{m^2}{g^2x}}\right),
\label{eq:freecondbulk}
\end{equation}
where $\mathrm{K}(u)$ is the complete elliptic integral of the first kind \cite{abramowitz}.
Note that by expanding this expression in the limit $x\to\infty$, the divergent logarithmic term $\frac{1}{2\pi}\frac{m}{g}\log x$ is indeed seen already in the free case.
In this way, we can extrapolate our lattice interacting condensate $\Sigma(m/g,x,N)$ first to infinite volume limit, $\Sigma(m/g,x)$, at a finite $x$ and a given $m/g$ and then subtract the infinite volume free condensate ($\Sigma_{\mathrm{free}}(m/g,x)$) given by Eq.~(\ref{eq:freecondbulk}):
\begin{equation}
\label{eq:subtr}
\Sigma_{\rm subtr}(m/g,x)=\Sigma(m/g,x)-\Sigma_{\mathrm{free}}(m/g,x),
\end{equation}
obtaining finally the subtracted condensate $\Sigma_{\rm subtr}(m/g,x)$, which can then be extrapolated to the continuum limit $x\rightarrow\infty$.
Note that a non-zero temperature does not bring any further divergence, hence the above renormalization scheme, subtracting the zero temperature free condensate in the infinite volume limit, can be applied for any $T$.
Actually, one can equivalently subtract the free condensate at any finite $T$.
This defines an alternative renormalization scheme
that we can also implement. Both options would lead to the correct value at $T=0$, i.e. compatible with the one directly obtained from the ground state 
calculations, but in order to compare to other results in the literature, we adopt in the following the $T=0$ renormalization scheme for all temperatures.

In the massless case, the temperature dependence of the chiral condensate was computed analytically by Sachs and Wipf~\cite{Sachs:1991en}:  
\begin{eqnarray}
   \left\langle {\bar \psi}\psi \right\rangle 
   &=& 
         \frac{m_{\gamma}}{2\pi} e^{\gamma} e^{2I(m_{\gamma}/T)} 
   = 
         \left\{ 
            \begin{array}{cl}  
                \frac{m_{\gamma}}{2\pi} e^{\gamma}  & \hspace{1cm} {\rm for} \hspace{1cm}  T\rightarrow 0  \\
                2T e^{-\pi T/m_{\gamma}}                   & \hspace{1cm} {\rm for} \hspace{1cm}  T\rightarrow \infty, 
            \end{array}  
         \right.
\end{eqnarray}
where $I(x) = \int_0^{\infty} \frac{dt}{1-e^{x\cosh(t)}}$, 
$\gamma\approx0.577216$ is the Euler-Mascheroni constant and $m_{\gamma}=g/\sqrt{\pi}$ is the non-perturbatively generated mass of the lowest lying boson (the vector boson).
According to the above formula, chiral symmetry is broken at any finite temperature (zero or non-zero) and it gets restored ($\Sigma=0$) only at infinite temperature. There is no phase transition, i.e. chiral symmetry restoration is smooth.

In the massive case, there is no analytical formula describing the temperature dependence of the condensate. However, the massive model was treated by Hosotani and Rodriguez with a generalized Hartree-Fock approach in Ref.~\cite{Hosotani:1998za}, yielding an approximate thermal dependence of $\Sigma$. In the following, we will confront our results with ones from this approximation and thus conclude about its validity.

\section{Tensor network approach}
\label{sec:TN}

In this work, we make use of two different applications of tensor network ansatzes. 
In order to obtain the results at zero temperature, we approximate 
variationally the ground state of the Schwinger model Hamiltonian  (\ref{eq:H})
on a finite lattice using a MPS.  
For the temperature dependence, we employ the matrix product operator (MPO) to describe the 
thermal equilibrium states at finite temperatures.

Although the details of these ansatzes and the basic algorithms involved can be found in the literature, for completeness
we compile in this section the fundamental ideas of both approaches, with special emphasis on the particularities associated to the problem at hand.

Given a system of $N$ sites, a MPS~\cite{vidal03eff,verstraete04dmrg,schollwoeck11age} is a state of the form
\begin{equation}
|\Psi\rangle = \sum_{i_0,\ldots i_{N-1}=0}^{d-1} \mathrm{tr}(A_0^{i_0}\ldots A_{N-1}^{i_{N-1}}) |i_0 \ldots i_{N-1}\rangle,
 \label{eq:MPS}
 \end{equation}
where $d$ is the dimension of the local Hilbert space for each site.
For two-level quantum systems, as in the case we are studying, $d=2$.
The state is parametrized by the $dN$ matrices, $A_k^{i}$, which have dimension $D\times D$, except for the ones at the edges,
$A_0^{i}$ and $A_{N-1}^{i}$, which, for the open boundary conditions we consider, are $D$-dimensional vectors.
The parameter $D$ is called the bond dimension, and determines the number of variational parameters in the ansatz.
The MPS can efficiently approximate ground states of local gapped Hamiltonians in one spatial dimension, and the 
ansatz lies at the basis of the success of the Density Matrix Renormalization Group (DMRG) 
method~\cite{white92dmrg,schollwoeck11age}.
In practice, they have been successfully applied to much more general problems, including long range 
interactions and two dimensional systems.

Different algorithms exist to find an MPS approximation to the ground state of a certain Hamiltonian. We use a variational search~\cite{verstraete04dmrg,schollwoeck11age}, in which the energy is minimized 
over the set of MPS with a given bond dimension,  $D$, by successively optimizing over one of the tensors, while keeping the rest fixed.
The procedure is repeated, while sweeping over all the tensors, until 
convergence is attained in the value of the energy, to a certain relative precision, $\varepsilon_{\mathrm{tol}}$,
ultimately limited by machine precision.
The computational cost of this procedure scales as $\mathcal{O}(d D^3)$ with the dimensions of the tensors.
The effect of running the algorithm with a limited bond dimension is to suffer a truncation error.
By running the algorithm with increasing values of $D$, we can estimate the magnitude of this error and extrapolate to the
$1/D \to 0$ limit, as discussed in detail in Sec. \ref{sec:results}.

Our previous works~\cite{Banuls:2013jaa,Banuls:2013zva} demonstrated that the MPS ansatz provides very good approximations 
to the ground state and lower lying excited states of the Schwinger model.

The MPS ansatz can be extended to the description of operators, and in particular density matrices~\cite{verstraete04mpdo,zwolak04mpo,pirvu10mpo}.
A matrix product operator (MPO) is thus of the form
\begin{eqnarray}
   \rho
   &=& 
         \sum_{\{i_k,j_k\}}{\rm Tr} \left( M[0]^{i_0j_0} \cdots M[N-1]^{i_{N-1}j_{N-1}} \right) |i_0\ldots i_{N-1}\rangle\langle j_0\ldots j_{N-1}|.
         \label{eq:MPO}
\end{eqnarray}
While any MPS (\ref{eq:MPS}) can represent a valid physical state, as far as it is normalized,
in order to describe a physical density operator, the MPO needs in addition to be positive. This condition cannot be guaranteed locally for 
generic tensors $M[k]$.
However, it is possible to ensure the positivity of a MPO using the \emph{purification ansatz}~\cite{verstraete04mpdo,delascuevas2013}, in which 
each tensor of the MPO has the form
$M[k]_{{\tilde \ell} {\tilde{r}}}^{ij}=\sum_p {A_{\ell' r'}^{ip}}^* A_{\ell r}^{jp}$.
This corresponds to a (pure state) MPS ansatz for an extended chain, with 
one ancillary system per site, such that $\rho$ is the reduced state for the original system, obtained by tracing out the ancillas.
It has been shown that thermal equilibrium states of local Hamiltonians can be well approximated by this kind of 
ansatz~\cite{hastings06gapped,molnar15gibbs} in arbitrary dimensions.

In the case of finite temperature, a MPO approximation can be constructed for the Gibbs state 
via imaginary time evolution of the identity operator~\cite{verstraete04mpdo}, 
$\rho(\beta)\propto e^{-\beta H}=e^{-\frac{\beta}{2} H} \Id e^{-\frac{\beta}{2} H}$, where $\beta\equiv1/T$ is the inverse temperature.
To achieve this, we apply a second order Suzuki-Trotter expansion \cite{trotter59,suzuki90} to the exponential,
and approximate every step of width $\delta=\beta/M$ by a product of five terms,
\begin{eqnarray}
e^{-\beta {H}}\approx
\left[ e^{- \frac{\delta}{2} {H}_e} e^{- \frac{\delta}{2} {H}_z} e^{-\delta {H}_o} 
       e^{- \frac{\delta}{2} {H}_z} e^{-\frac{\delta}{2} {H}_e} \right]^M,
\label{eq:trotter}
\end{eqnarray}
where $H_z=H_m+H_g$ is diagonal in the $z$ basis, and the hopping term is split in two sums $H_{hop}=H_e+H_o$,
with the $H_{e}$ ($H_{o}$) term containing the two-body terms that act on each even-odd (odd-even) pair of sites.
If each of the exponential terms can be exactly computed, the error of this approximation scales as $O(\delta^2)$.
The exponentials of $H_e$ and $H_o$ have indeed an exact MPO expression with constant bond dimension $4$.
The term $H_z$ contains long range interactions, but its structure allows us to also write it exactly as a MPO, 
with bond dimension $(N+1)$,
as detailed in Ref.~\cite{Banuls:2015sta}.
The only non-vanishing elements of the tensors specifying the MPO are
$(M_n^{ii})_{L_{n-1} L_n}=e^{-\delta h_n}$,
for $L_{n}=L_{n-1}+\frac{1}{2}[(-1)^n+(\sigma_n^z)_{ii}]$, where 
$h_n=\frac{\mu}{2} \left[ 1+ (-1)^n \sigma_n^z \right] +L_n^2$  for $n<N-1$,
and $h_{N-1}=\frac{\mu}{2} \left[ 1+ (-1)^{N-1} \sigma_{N-1}^z \right]$.
The virtual bond then carries the information about the electric flux on each link,
which can assume values $L_n\in[-N/2,N/2]$.
Instead of working with the exact exponential of $H_z$, which has a bond dimension $N+1$, we find it convenient, given the large system sizes 
we want to study, to 
truncate the dimension of the MPO, by defining a maximum value the virtual bond can attain, $|L_n| \leq L_{\mathrm{cut}}$.
This is equivalent to truncating the physical space to those states where the electric flux on a link cannot exceed $L_{\mathrm{cut}}$
and is thus related to approaches where one explicitly truncates the maximum allowed occupation number of the bosonic gauge 
degrees of freedom~\cite{Buyens:2013yza}.

Starting with the identity operator, $\rho(0)$, which has a trivial expression as a MPO with bond dimension one, we successively apply steps
of the evolution, using the approximation above, and approximate the result by a MPO with the desired maximum bond dimension. 
This is achieved with the help of a Choi isomorphism~\cite{choi}, $|i\rangle\langle j| \rightarrow |i\rangle\otimes |j\rangle$,
to vectorize the density operators, such that the MPO is transformed in a MPS, {with physical dimension per site $d^2$}, on which the evolution steps act linearly.
The approximated effect of the evolution is then found by minimizing the Euclidean distance between the original and final MPS.
The procedure can be repeated until inverse temperature $\beta/2$ is reached. Then we construct $\rho(\beta)\propto\rho(\beta/2)^{\dagger} \rho(\beta/2)$
(up to normalization) such that the purification ansatz is realized and we ensure a positive thermal equilibrium state.
 The computational cost of this calculation is the same as that of time evolution of a MPS state, with the increased physical dimension, i.e. it scales as $\mathcal{O}(d^2D^3)$.

Using the MPO ansatz with limited bond dimension induces also a truncation error in the $T>0$ case, which is not equivalent to the one 
described for $T=0$.
First of all, different ansatzes are used for both cases, and while the MPS truncation in the pure state case can be related to the entanglement in the state, 
the same is not true for the MPO ansatz in the case of mixed states. \footnote{In the case of operators one should instead talk about operator space entanglement entropy, 
a measure related to truncation error in the MPO that was introduced in Ref.~\cite{PhysRevA.76.032316}.}
Moreover, the distinct numerical algorithms used in both cases also mean that errors are introduced in different ways.
In the thermal algorithm, each application of one of the exponential factors in (\ref{eq:trotter}) potentially increases the bond dimension of the resulting MPO.
Hence, after every step, the ansatz needs to be truncated to the maximum desired value of the bond dimension. In practice, this is achieved 
by minimizing a cost function that corresponds to the Frobenius norm of the difference to the true evolved operator.
As in the ground state search, this optimization is done by an alternating least squares (ALS) scheme, in which all tensors but one are fixed, 
and repeated sweeping is performed over the chain.
Also in this case, we use a tolerance parameter, $\varepsilon_{\mathrm{tol}}$, to decide about the convergence of the iteration,
but now the value bounds the relative change in the cost function during the sweeping that follows 
the application of each single exponential factor.
This procedure leads to errors accumulating along the thermal evolution, 
and while at $\beta=0$ the state can be exactly written as a MPO with $D=1$,
the largest truncation errors will occur for the lowest temperatures.
Thus, recovering zero temperature results from such a procedure is a non-trivial check that the method is working correctly.
The $T=0$ calculation, in contrast, does not suffer from this effect, as it directly targets the ground state variationally.

Additionally, the Suzuki-Trotter expansion (\ref{eq:trotter}) introduces another systematic error in the thermal evolution, by using a finite step width $\delta$, which we need to extrapolate to $\delta \to 0$,
and another one in the form of the truncation of the physical subspace to a maximum $L_{\mathrm{cut}}$, described above.
All these factors need to be taken into account when performing the extrapolations required to extract the continuum values of the observables under study  (see Sec. \ref{sec:results} for details).

\section{Results}
\label{sec:results}
\subsection{Zero temperature}
\label{sec:zero}
We begin with our results for the ground state chiral condensate for various fermion masses. For the massless case, an analytical result can be obtained, $\Sigma/g=e^\gamma/2\pi^{3/2}\approx0.159929$. 
We are able to reproduce this number with great accuracy and also obtain results in the massive case, where no analytical results exist.

Our numerical procedure consists in computing several sets of data points corresponding to different values of ($D$, $N$, $x$) and extrapolating in the way described below.
\\
\textbf{Infinite bond dimension ($D\rightarrow\infty$) extrapolation.}
We use several values of $D\in[40,160]$ to check the effects from changing the bond dimension. Our final value is taken as the condensate corresponding to the largest computed value of $D=160$ and its error as the difference between the value for $D=160$ and $D=140$. The lower values of $D$ serve to ensure that the two highest bond dimensions are large enough, such that it can be argued that the difference between $D=\infty$ and $D=160$ is smaller than the one between $D=160$ and $D=140$, which makes our error estimate valid.

\begin{figure}[t!]
\begin{center}
\includegraphics[width=0.345\textwidth,angle=270]{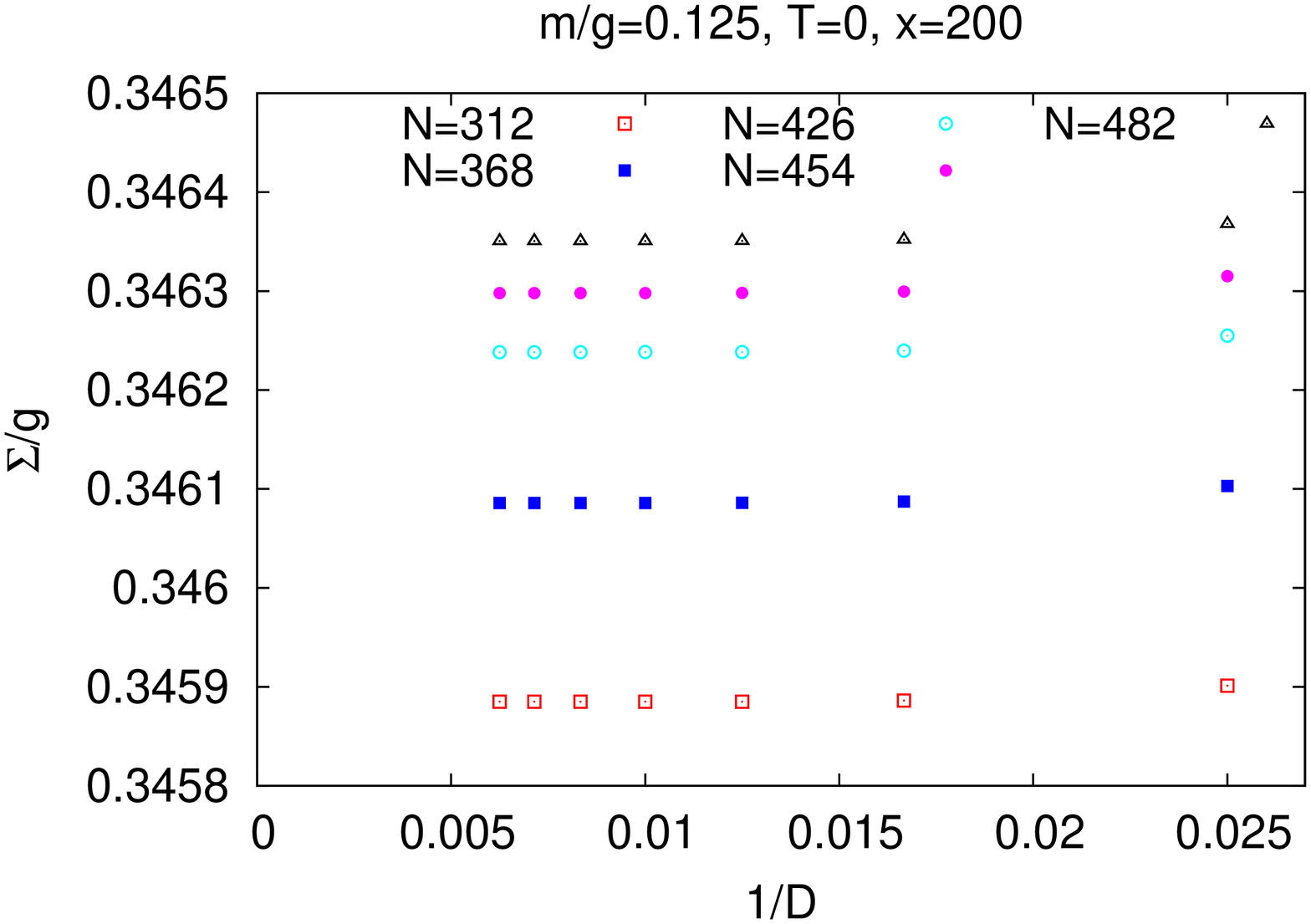}
\includegraphics[width=0.345\textwidth,angle=270]{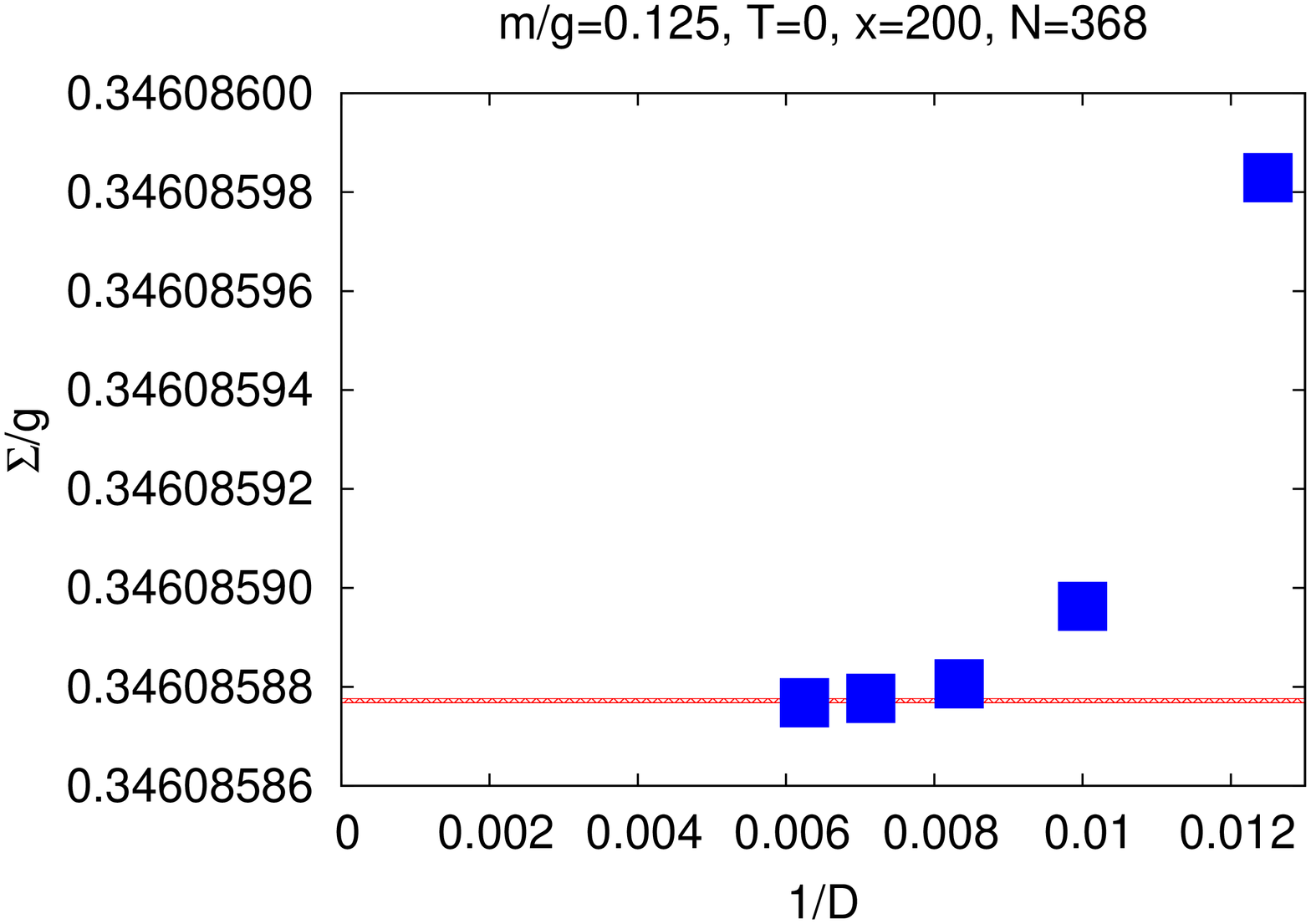}
\caption{\label{fig:D200}Examples of the $D$-dependence of the ground state chiral condensate for $m/g=0.125$, $x=200$ and five system sizes (left). The right plot shows a zoom into the region $D\in[80,160]$ for $N=368$. The red band represents the uncertainty related to the bond dimension, taken as explained in the text.}
\end{center}
\end{figure}

\begin{figure}[t!]
\begin{center}
\includegraphics[width=0.345\textwidth,angle=270]{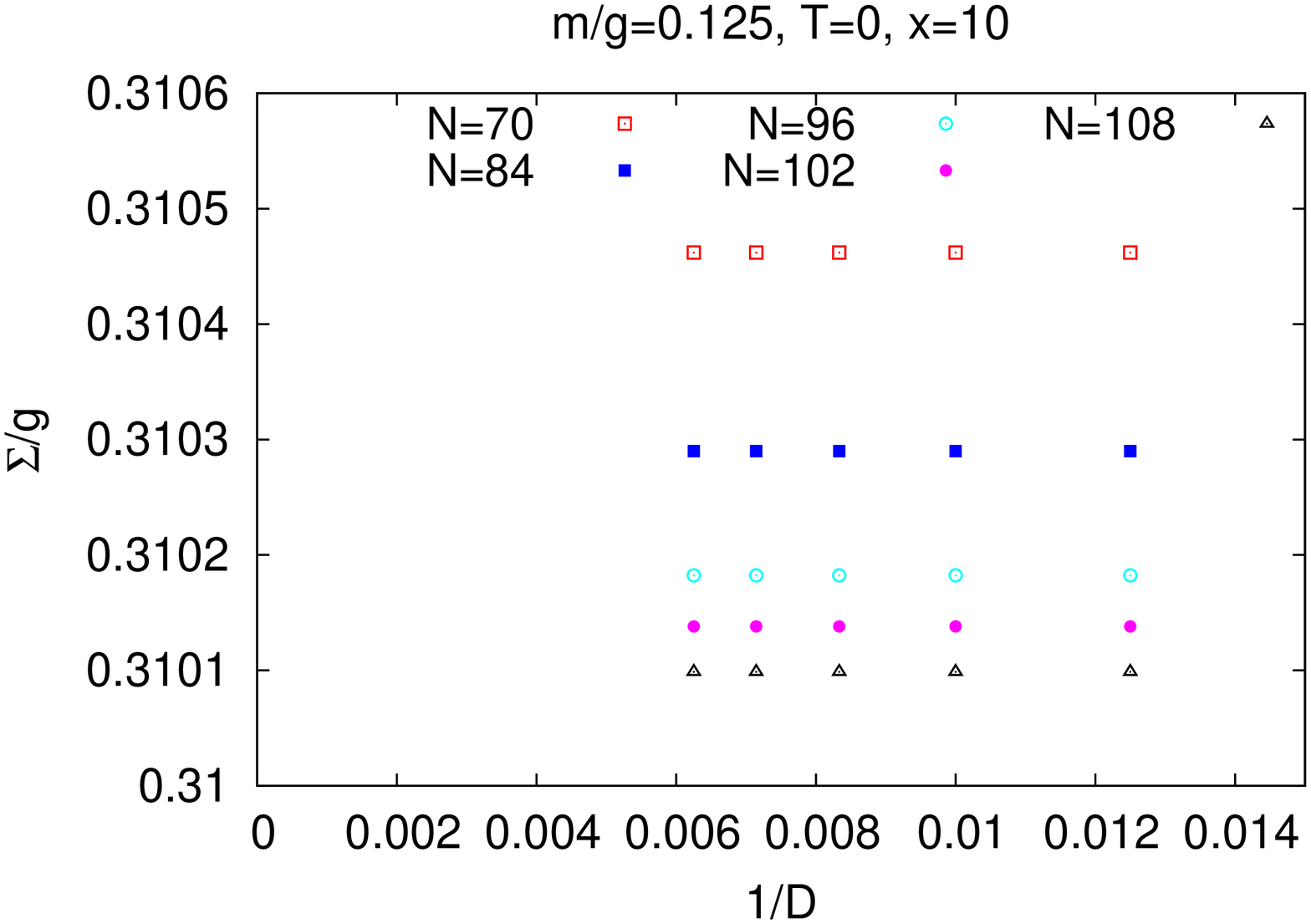}
\includegraphics[width=0.345\textwidth,angle=270]{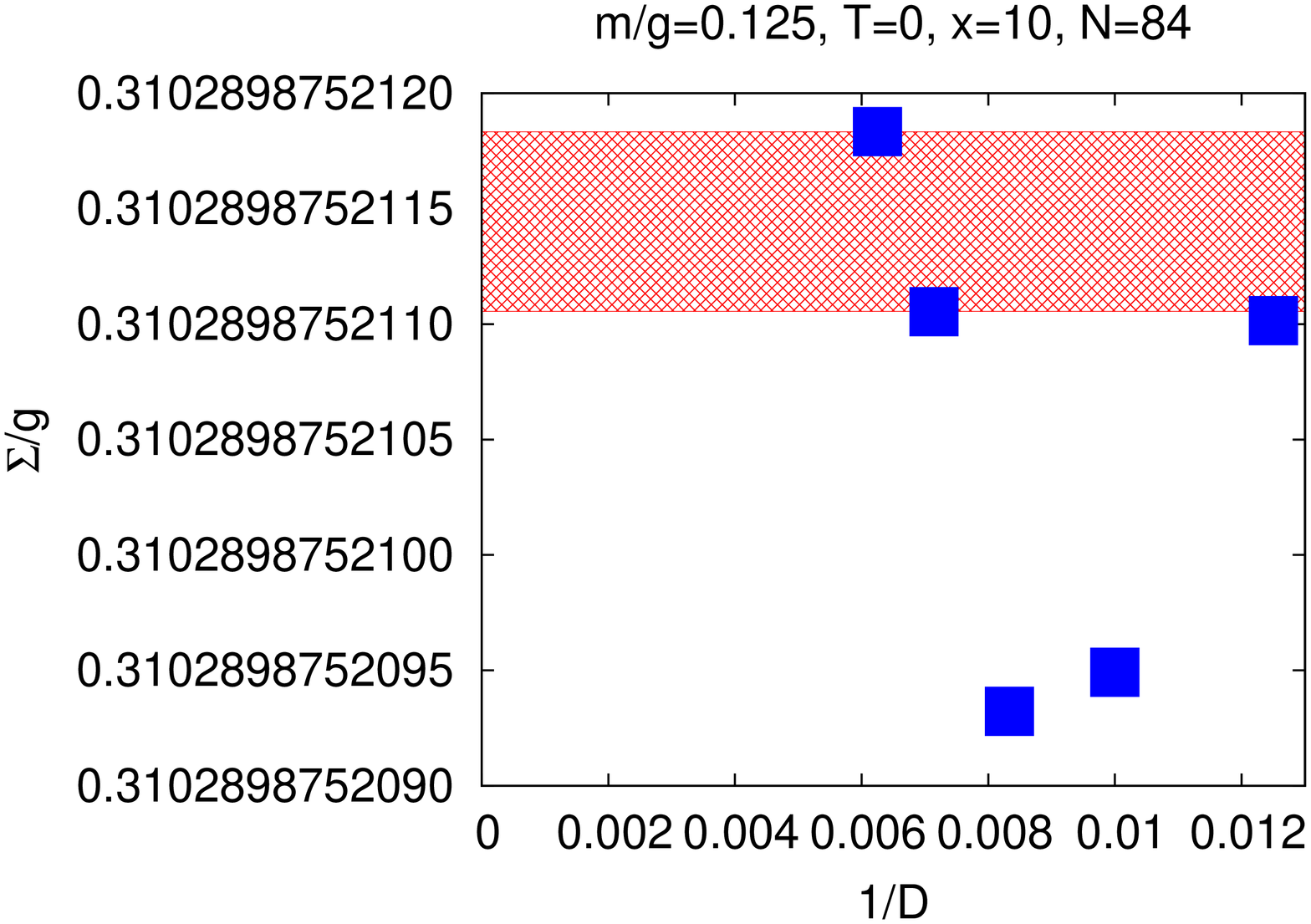}
\caption{\label{fig:D10}Examples of the $D$-dependence of the ground state chiral condensate for $m/g=0.125$, $x=10$ and five system sizes (left). The right plot shows a zoom into the region $D\in[80,160]$ for $N=84$. See comments in the text about the irregular approach to the $1/D=0$ limit. The red band represents the uncertainty related to the bond dimension, taken as explained in the text.
}
\end{center}
\end{figure}

A typical example of such extrapolation is shown in Fig.~\ref{fig:D200} for $x=200$ and in Fig.~\ref{fig:D10} for $x=10$, at $m/g=0.125$.
In both cases, we observe very good convergence towards the $1/D=0$ limit, with the above defined error from this step being of $\mathcal{O}(10^{-9})$ for the former and $\mathcal{O}(10^{-12})$ for the latter.
This error is represented by a red band.
Note that despite going to $D=160$, the convergence in bond dimension is so good that actually even with $D=40$ we would already obtain the result with an outstanding precision, of $\mathcal{O}(10^{-8})$ for $x=200$ (i.e. only an order of magnitude worse than with $D=160$) or even of $\mathcal{O}(10^{-12})$ for $x=10$ (i.e. the same as with $D=160$).
The $x=10$ case illustrates that in some cases the convergence in $D$ is so good that our uncertainty comes from issues with the numerical precision.
The MPS optimization procedure is considered to be converged when the relative change in the ground state energy in subsequent sweeps falls below a certain tolerance parameter, taken to be $\varepsilon_{\mathrm{tol}}=10^{-12}$ in our case.
Notice, however, that this precision refers to the ground state energy, which typically converges better than other observables, so it will correspond to a somewhat worse precision in the chiral condensate, which we estimate to be in the $10^{-11}-10^{-10}$ region.
In the $x=10$ case, the variation of $\Sigma/g$ values for different $D$ becomes smaller than this, which explains the irregular behaviour of the $D$-dependence for this case (left plot of Fig.~\ref{fig:D10}), compared to the apparently regular convergence for the case $x=200$.
We account for this bias (that happens only for our smallest $x$ values) in our next step, the infinite volume extrapolation.
We emphasize that this is definitely not a drawback of the method, but even better precision could be attained for certain parameter ranges with the same $D$ values, by 
adopting a more demanding convergence criterion.
On the other hand, since the ultimate limit of machine precision, which we label by $\epsilon_{\mathrm{num}}$, affects the optimization of individual tensors, so that after one sweep over the whole chain, it may affect the value of the energy
in $\mathcal{O}(N \epsilon_{\mathrm{num}}).$ This means that for chains of hundreds of sites, as required for the largest values of $x$ we explore, 
$\varepsilon_{\mathrm{tol}}=10^{-12}$ is the best allowed by double precision numerics.

\begin{figure}[t!]
\begin{center}
\includegraphics[width=0.345\textwidth,angle=270]{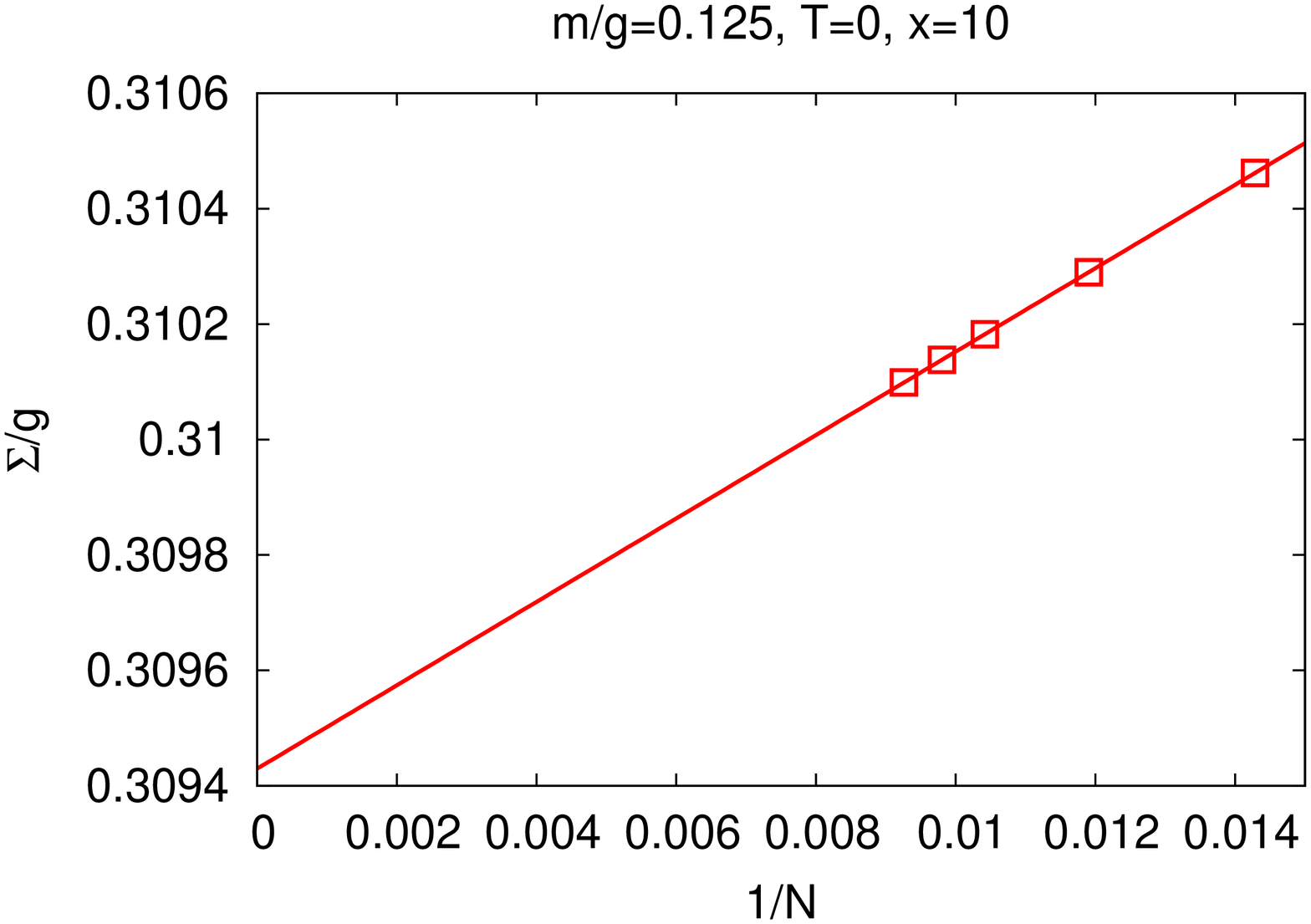}
\includegraphics[width=0.345\textwidth,angle=270]{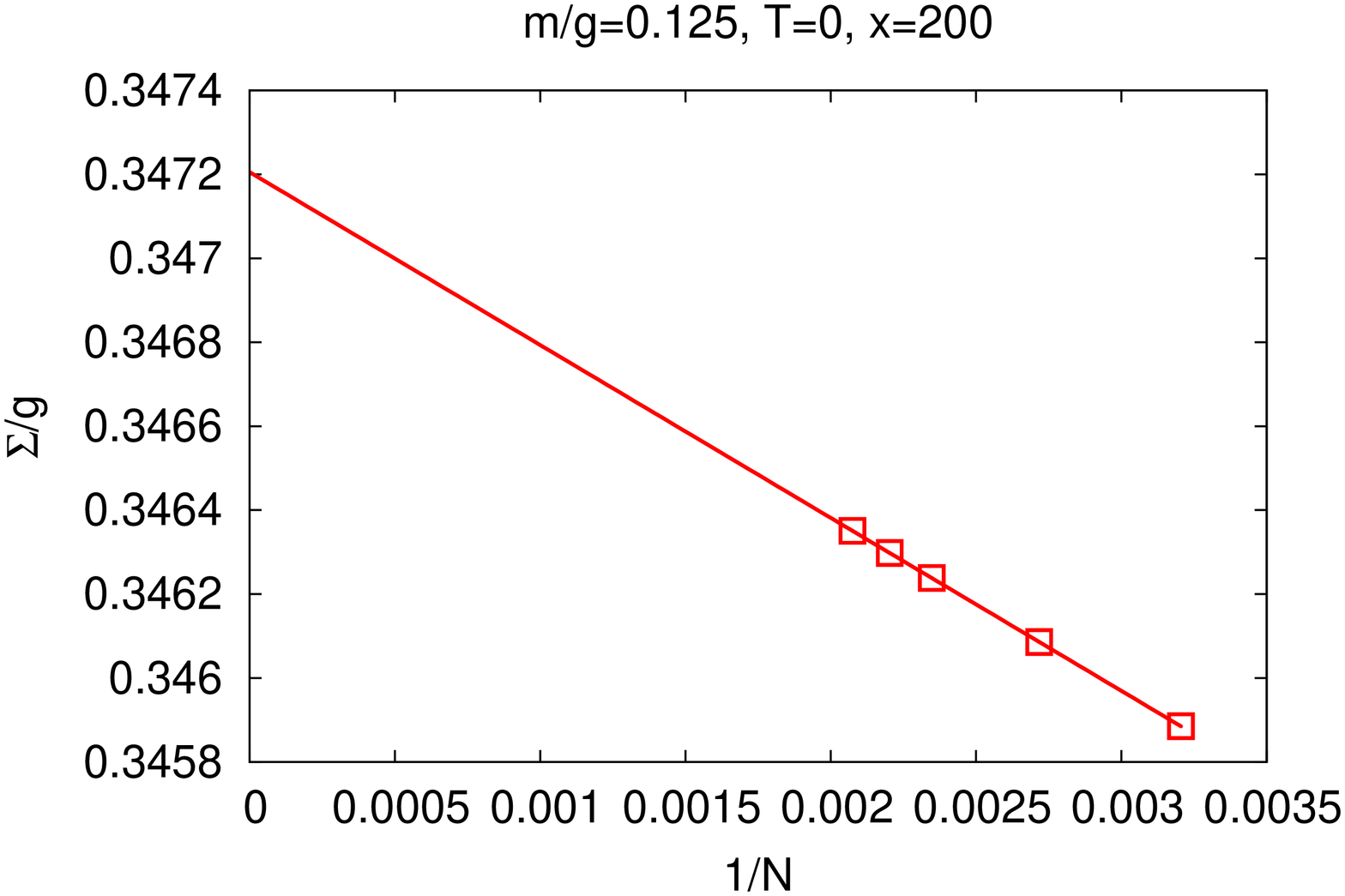}
\caption{\label{fig:N0}Examples of the infinite volume extrapolations of the $T=0$ chiral condensate for $x=10$ (left) and $x=200$, at $m/g=0.125$. Lines are fits of Eq.~(\ref{eq:fitN}). The points have error bars, but they are too small to be seen.
}
\end{center}
\end{figure}

\noindent\textbf{Infinite volume ($N\rightarrow\infty$) extrapolation.}
The results corresponding to our estimates of the $D\rightarrow\infty$ limit can then be extrapolated to infinite volume by using a linear fitting ansatz:
\begin{equation}
\label{eq:fitN}
\Sigma(m/g,x,N)=\Sigma(m/g,x)+\frac{\alpha(m/g,x)}{N},
\end{equation}
where $\Sigma(m/g,x,N)$ is the infinite-$D$ condensate for a fixed fermion mass, volume and lattice spacing. The fitting parameters are $\Sigma(m/g,x)$ (infinite volume condensate at a given lattice spacing and fermion mass) and the mass and lattice spacing-dependent slope of the finite volume $1/N$ correction, $\alpha(m/g,x)$.
We show an example of such extrapolation in Fig.~\ref{fig:N0}, again for $x=10$ (left) and $x=200$ (right), at $m/g=0.125$.
We always choose the volumes to be large enough, such that the above linear fitting ansatz yields a good description of data.
We have found that this holds when the volumes used are scaled proportionally to $\sqrt{x}$ and we take $N=\{22\sqrt{x},\,26\sqrt{x},\,30\sqrt{x},\,32\sqrt{x},\,34\sqrt{x}\}$. 
Indeed, in all cases where no issues with machine precision are observed, this leads to very good fits.
The resulting error of the fitting coefficient $\Sigma(m/g,x)$ is the propagated error from the $D$-extrapolation.
For very small values of $x$ (lower than approx. 30), we need to deal with the numerical precision bias. 
The errors from the $D$-extrapolation are in such case underestimated, since they do not take into account the finite numerical precision.
This leads to $\chi^2/{\rm dof}$ values of $\mathcal{O}(10-100)$.
However, we know from the analysis for large values of $x$ that the linear fitting ansatz (\ref{eq:fitN}) yields an excellent description of data, with $\chi^2/{\rm dof}$ usually much smaller than 1.
Hence, we account for the bias by inflating the $D$-extrapolation errors to such levels that $\chi^2/{\rm dof}=1$ by construction.
In this way, the final error after the infinite volume extrapolation step is properly rescaled and becomes comparable to the one at larger $x$ (e.g. approx. $8.2\cdot10^{-10}$ for $x=10$ and $6.3\cdot10^{-9}$ for $x=200$).
In the end, all our errors of infinite volume condensates, $\Sigma(m/g,x)$, differ by less than an order of magnitude in the whole considered range of $x$ and for all fermion masses.
\\
\textbf{Continuum limit ($x\rightarrow\infty$) extrapolation.}
Finally, the infinite volume results $\Sigma(m/g,x)$ can be extrapolated to the continuum limit.
First, we subtract the infinite volume free condensate according to Eq.~(\ref{eq:subtr}), obtaining the subtracted condensate $\Sigma_{\rm subtr}(m/g,x)$.
Then, we apply the following fitting ansatz:
\begin{equation}
\label{eq:cont} 
\Sigma_{\rm subtr}(m/g,x)=\Sigma_{\rm subtr}(m/g)+\frac{a(m/g)}{\sqrt{x}}\log(x)+\frac{b(m/g)}{\sqrt{x}}+\frac{c(m/g)}{x},
\end{equation}
with fitting parameters $\Sigma_{\rm subtr}(m/g)$ (the continuum condensate for a given fermion mass), $a(m/g)$, $b(m/g)$ and $c(m/g)$.
This is a fitting ansatz quadratic in the lattice spacing (the role of the lattice spacing is played by $1/\sqrt{x}$), with logarithmic corrections.
The latter appear already in the free theory, where their presence can be shown analytically (see Sec.~\ref{sec:schwinger}).
Note that the final result obtained from this procedure will, to some extent, depend on the fitting range in $1/\sqrt{x}$.
To quote final values independent from such choices, we adopt a systematic procedure analogous to the one we used in our spectrum investigation in the Schwinger model, described in detail in the appendix of Ref.~\cite{Banuls:2013jaa}.
In short, this consists in performing fits in different possible fitting ranges by varying the minimal and maximal values of $x$ entering the fits.
The number of fits that we obtain in this way is of $\mathcal{O}(100)$ and allows us to build a distribution of the continuum values, weighted with $\exp(-\chi^2/{\rm dof})$ of the fits.
The final value that we quote is the median of the distribution and the systematic error from the choice of the fitting range comes from the 68.3\% confidence interval (such that in the limit of infinite number of fits it corresponds to the width of a resulting Gaussian distribution).
This error is then combined in quadrature with our propagated error from $D$- and $N$-extrapolations, which we take as the error of one selected fit, taken to be the one in the interval $x\in[20,600]$.

\begin{figure}[t!]
\begin{center}
\includegraphics[width=0.345\textwidth,angle=270]{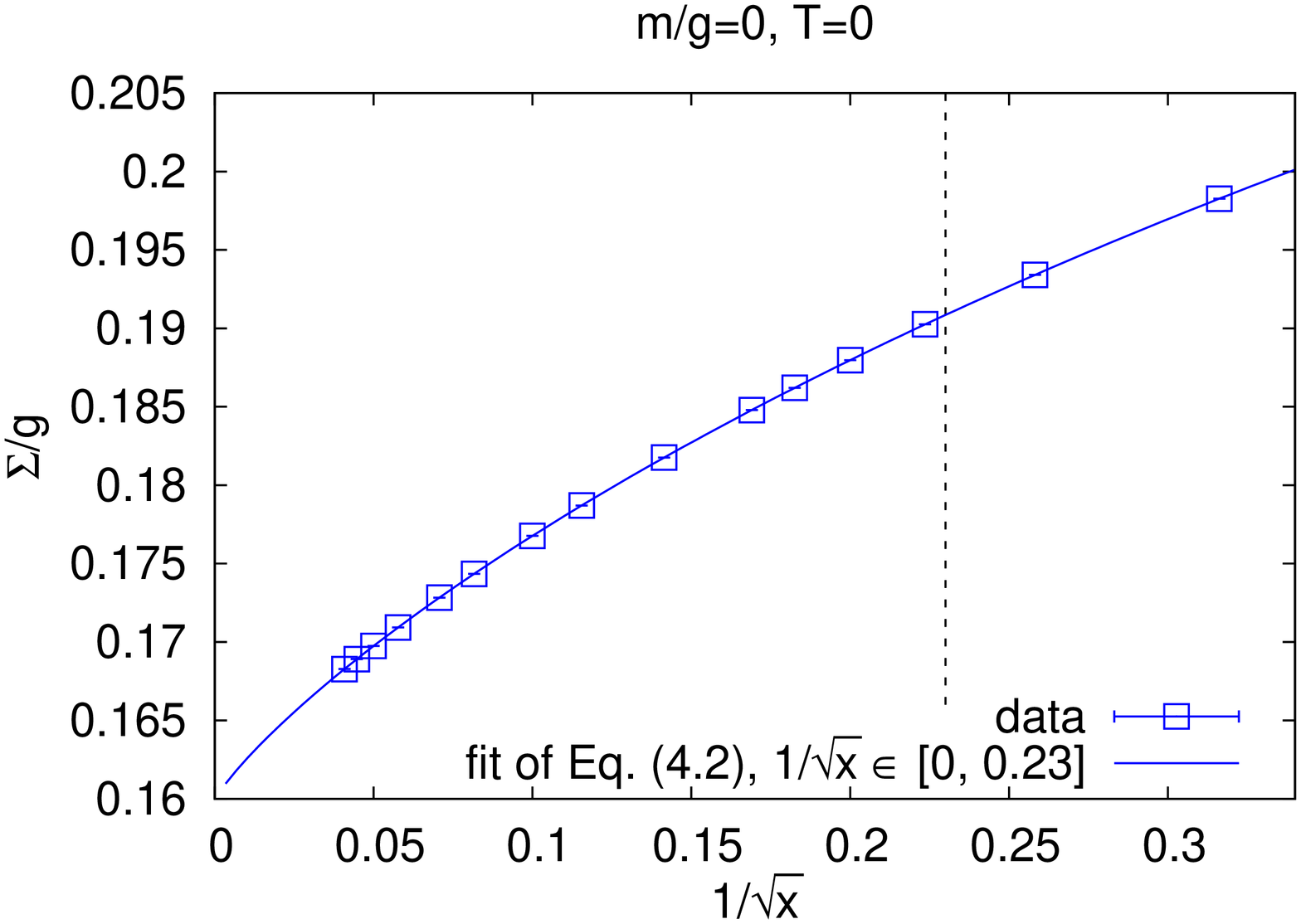}
\includegraphics[width=0.345\textwidth,angle=270]{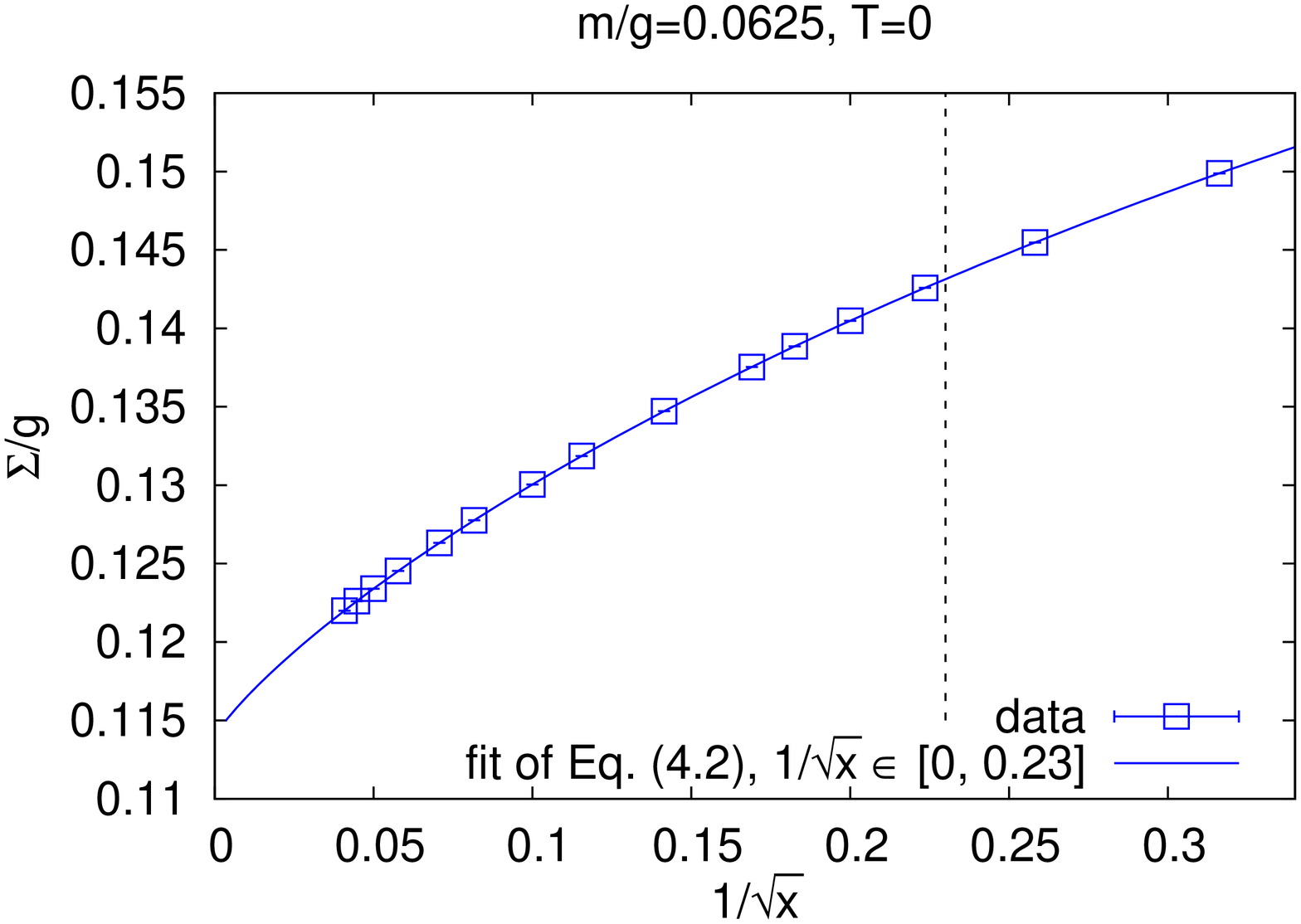}
\includegraphics[width=0.345\textwidth,angle=270]{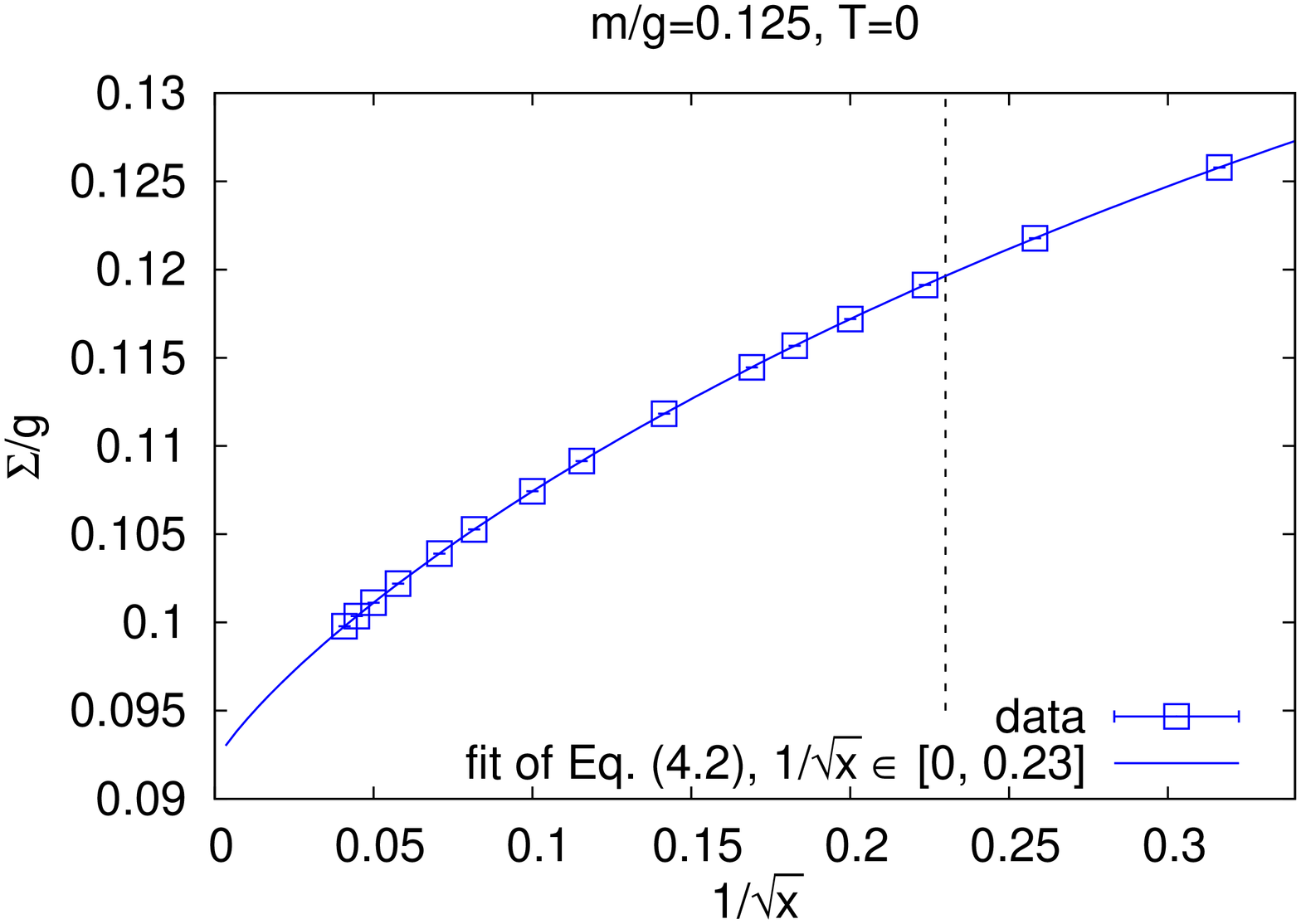}
\includegraphics[width=0.345\textwidth,angle=270]{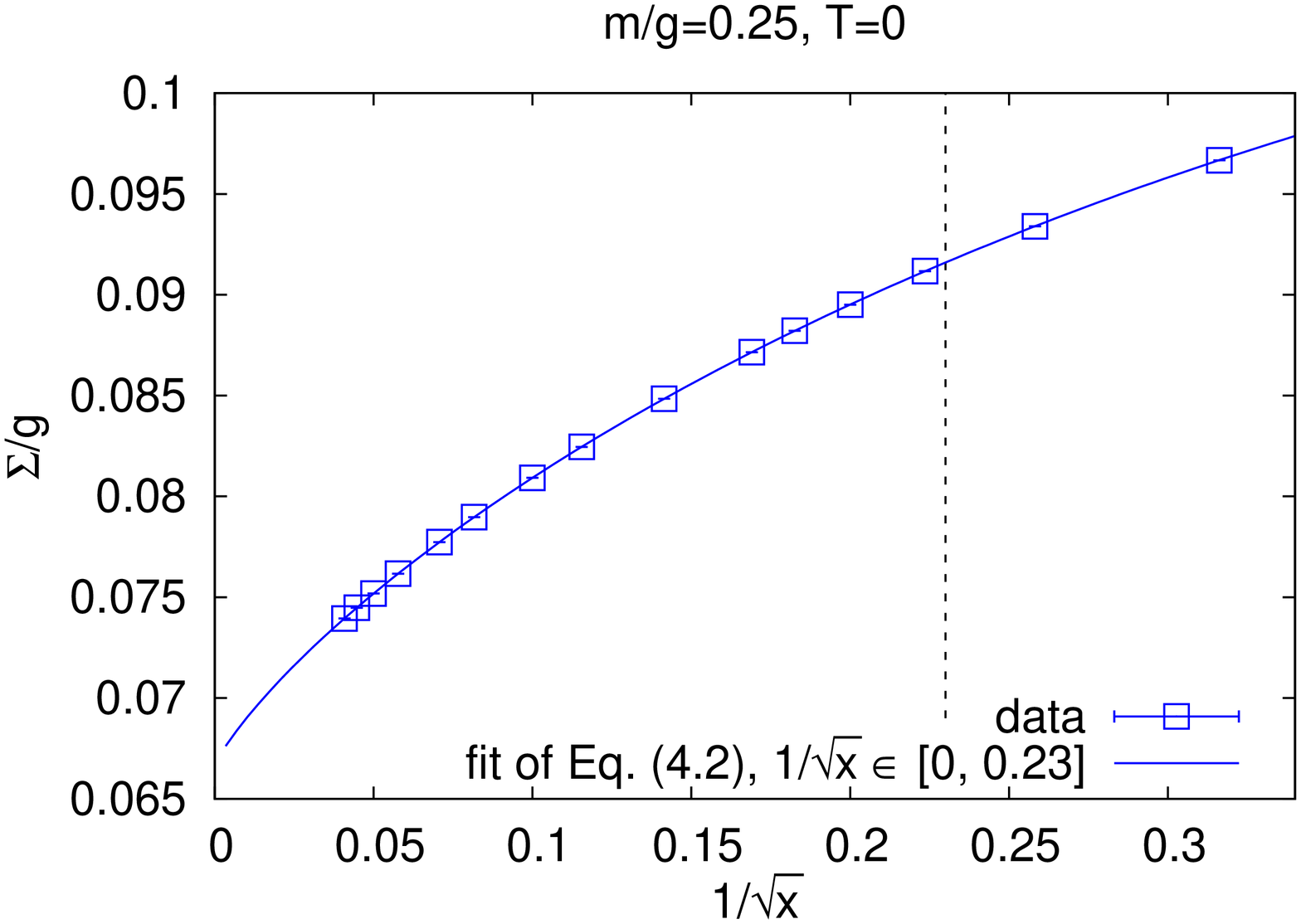}
\includegraphics[width=0.345\textwidth,angle=270]{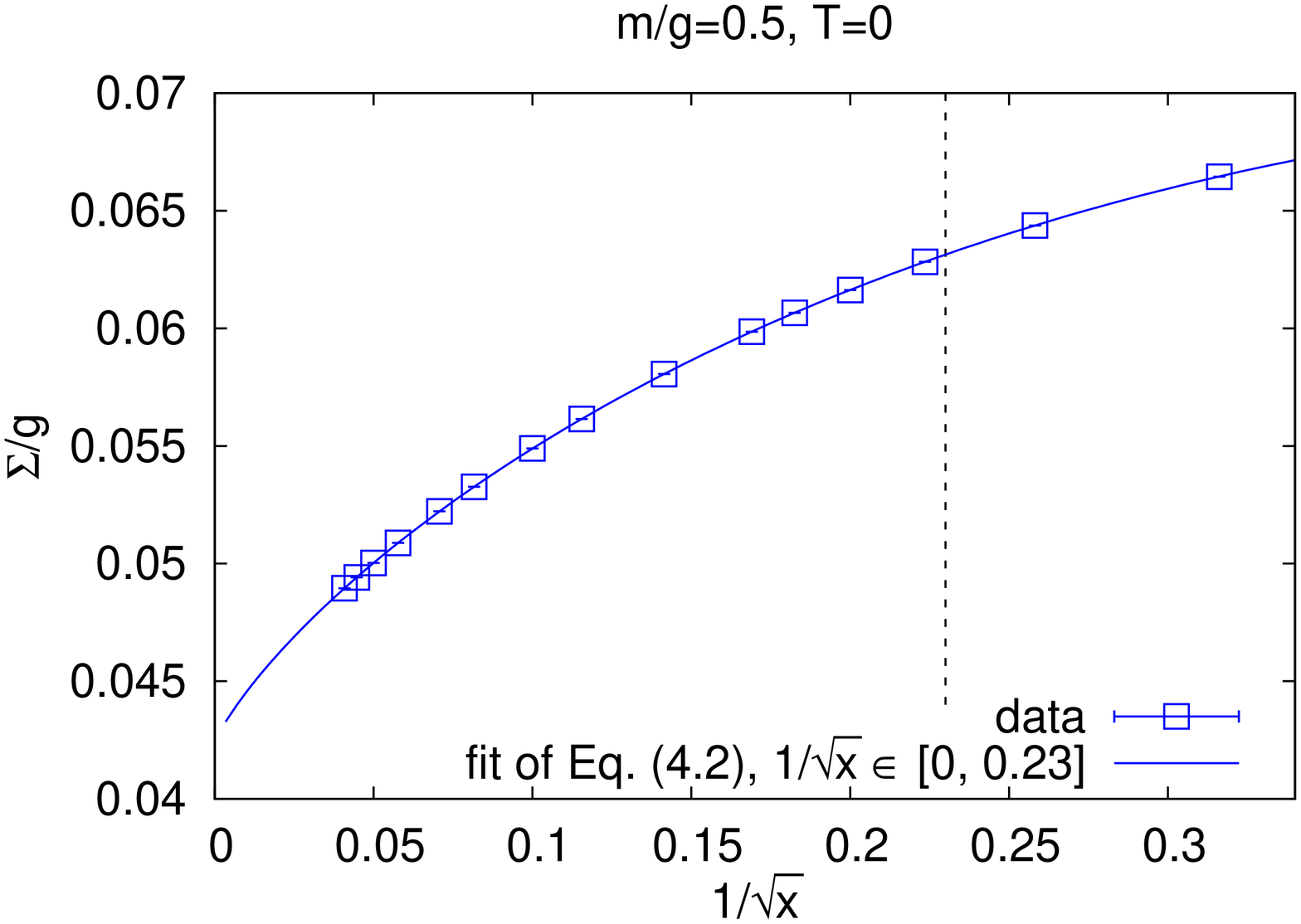}
\includegraphics[width=0.345\textwidth,angle=270]{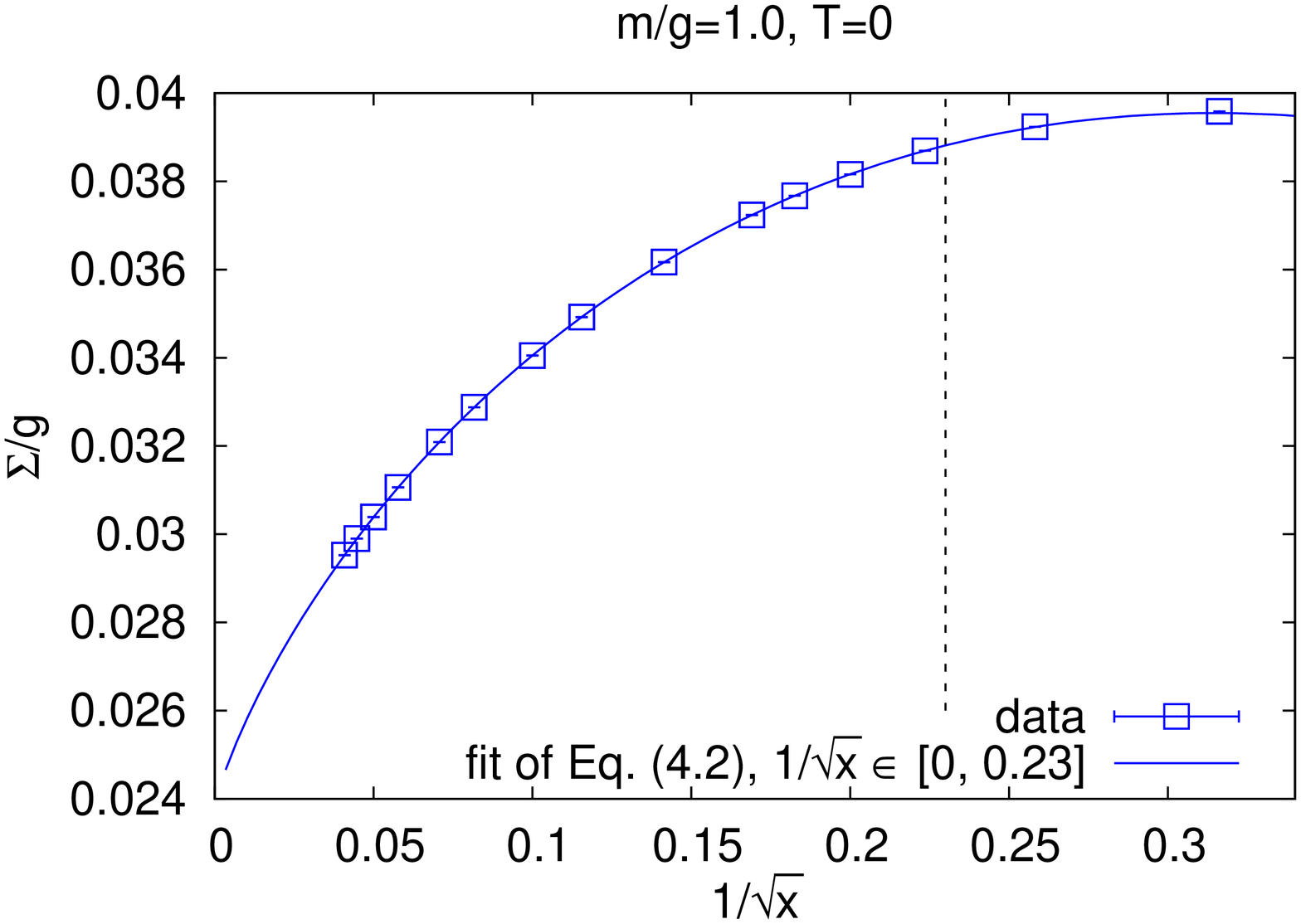}
\caption{\label{fig:cont0}Continuum limit extrapolations of the $T=0$ chiral condensate for all our fermion masses. Shown are the data for the subtracted infinite volume condensate and the fits of Eq.~(\ref{eq:cont}) in the interval $1/\sqrt{x}\in[0,0.23]$ ($x\in[20,600]$, i.e. left of the dashed line). These fits are example fits that enter our procedure of extraction of the systematic uncertainty related to the choice of the fitting range. Note that the rightmost two points are not included in the fit, but are still rather well described by it.
}
\end{center}
\end{figure}
\begin{table}[t!]
\begin{center}
\begin{tabular}{|c|c|c|c|}
\hline
&
\multicolumn{3}{|c|}{Subtracted condensate} \\
\hline
\multirow{2}*{$m/g$}  & \multirow{2}*{Our result} & \multirow{2}*{Ref.~\cite{Buyens:2014pga}} & Exact ($m=0$) \\
&&& or Ref.~\cite{Hosotani:1998za} ($m>0$)\\
\hline
0 & 0.159929(7) & 0.159929(1) & 0.159929 \\
\hline
0.0625 & 0.1139657(8) & -- & 0.1314 \\
\hline
0.125 & 0.0920205(5) & 0.092019(2) & 0.1088 \\
\hline
0.25 &  0.0666457(3) & 0.066647(4) & 0.0775 \\
\hline
0.5 & 0.0423492(20) & 0.042349(2) & 0.0464 \\
\hline
1.0 & 0.0238535(28) & 0.023851(8) & 0.0247\\
\hline
\end{tabular}
\caption{\label{tab:T0} Final continuum values of the $T=0$ chiral condensate (in units of $g$) for the used fermion masses. We compare with results from Ref.~\cite{Buyens:2014pga} and with the analytical result in the massless case or the approximated result from Ref.~\cite{Hosotani:1998za} in the massive case.}
\end{center}
\end{table}

Our continuum limit extrapolations are shown in Fig.~\ref{fig:cont0} for all fermion masses that we considered. We show in these plots the fit from which we estimated our propagated error from earlier extrapolations ($x\in[20,600]$), i.e. one of the fits that enter the distribution built to assess our final values and their uncertainties. The final values for each fermion mass are summarized in Tab.~\ref{tab:T0}.
We compare to the result of a similar calculation in Ref.~\cite{Buyens:2014pga} and to the exact result in the massless case or the approximation of Ref.~\cite{Hosotani:1998za}.
For the former, we observe perfect agreement, which is quite remarkable given the precision of both results being at the $\mathcal{O}(0.01\%-0.001\%)$ level.
Similarly good is the agreement with the analytical result at $m=0$. 
We will comment more on the agreement with Ref.~\cite{Hosotani:1998za} in the next subsection.

\subsection{Thermal evolution}
\label{sec:thermal}
In our previous papers \cite{Saito:2014bda,Saito:2015ryj,Banuls:2015sta}, we showed results for the temperature dependence of the chiral condensate in the massless case.
We employed a method without any truncations in the gauge sector and found that it is numerically very demanding to achieve lattice spacings small enough to reliably extrapolate to the continuum at high temperatures.
This led us to the method of introducing a finite cut-off, $L_{\rm cut}$, in the gauge sector and we showed that this method works very well in the massless case, allowing for good precision of results for the whole range of temperatures.
In the present paper, we test the method, explained in detail in Sec.~\ref{sec:TN}, in the massive case.
Although this method is different from the one used for $T=0$, the analysis procedure at a given temperature is rather similar to the one described in the previous subsection.
We begin by shortly outlying the new parts of the analysis in the thermal case.
In the following, we typically express the temperature with its inverse, $\beta\equiv1/T$.

\begin{figure}[t!]
\begin{center}
\includegraphics[width=0.345\textwidth,angle=270]{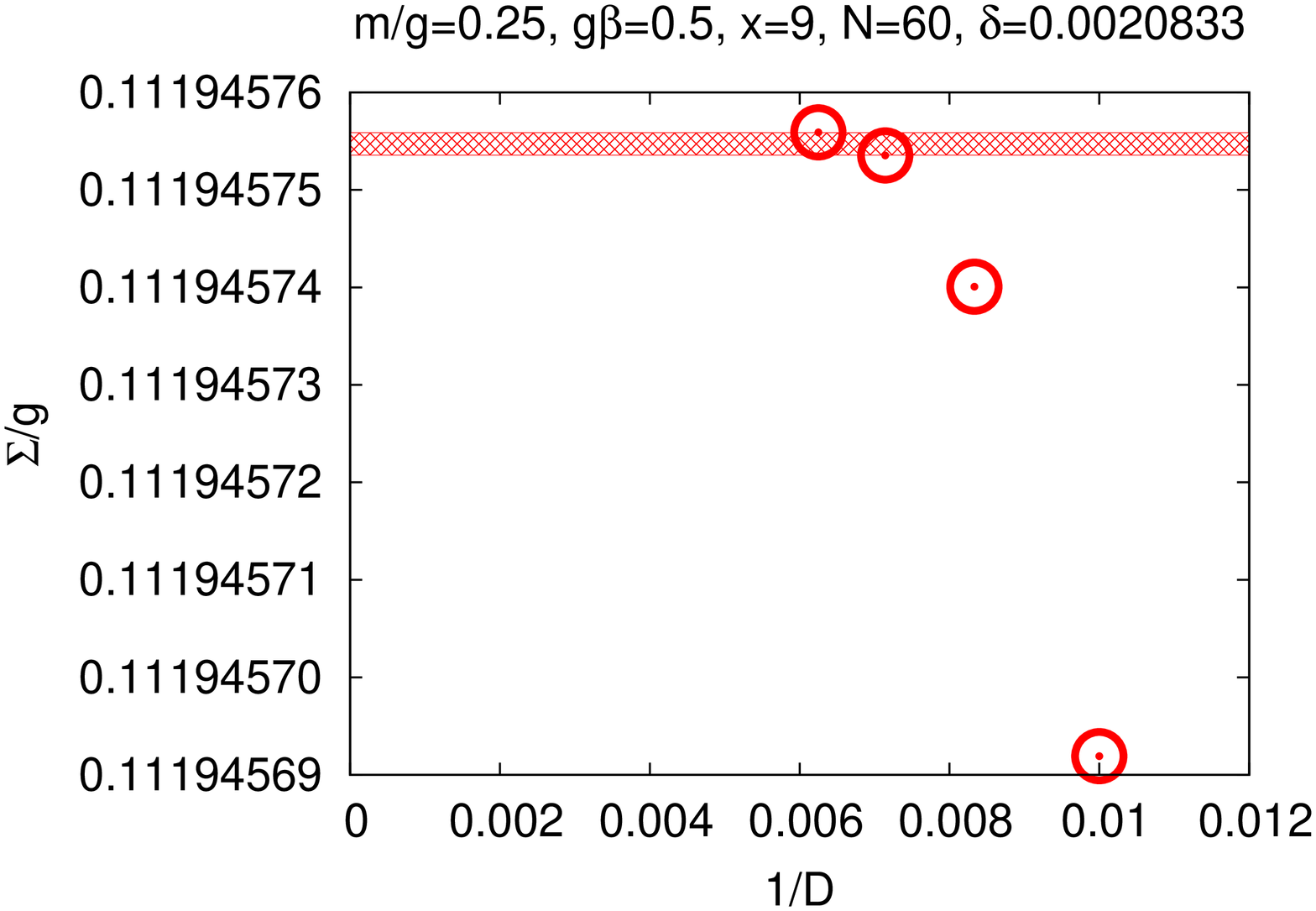}
\includegraphics[width=0.345\textwidth,angle=270]{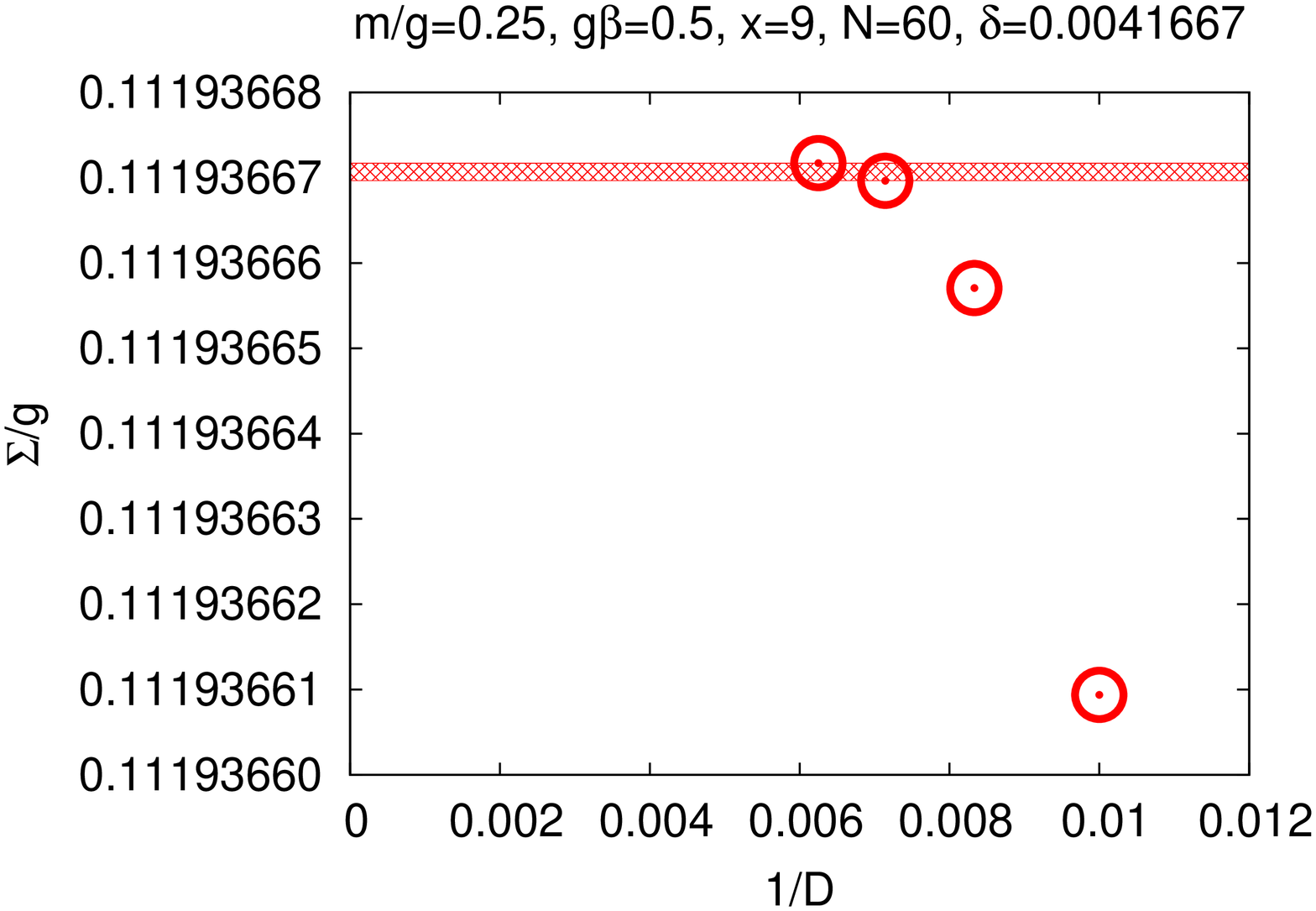}
\includegraphics[width=0.345\textwidth,angle=270]{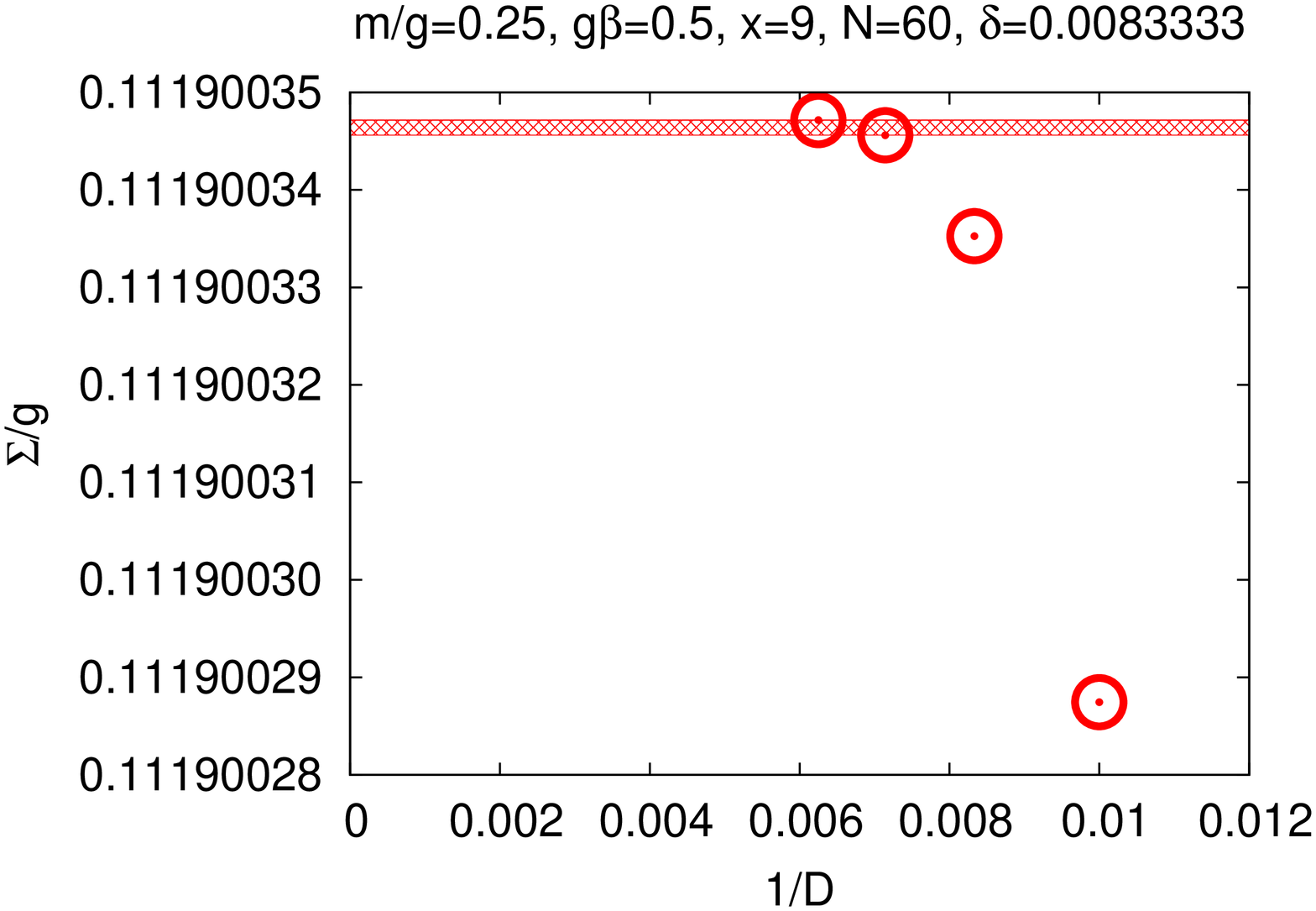}
\includegraphics[width=0.345\textwidth,angle=270]{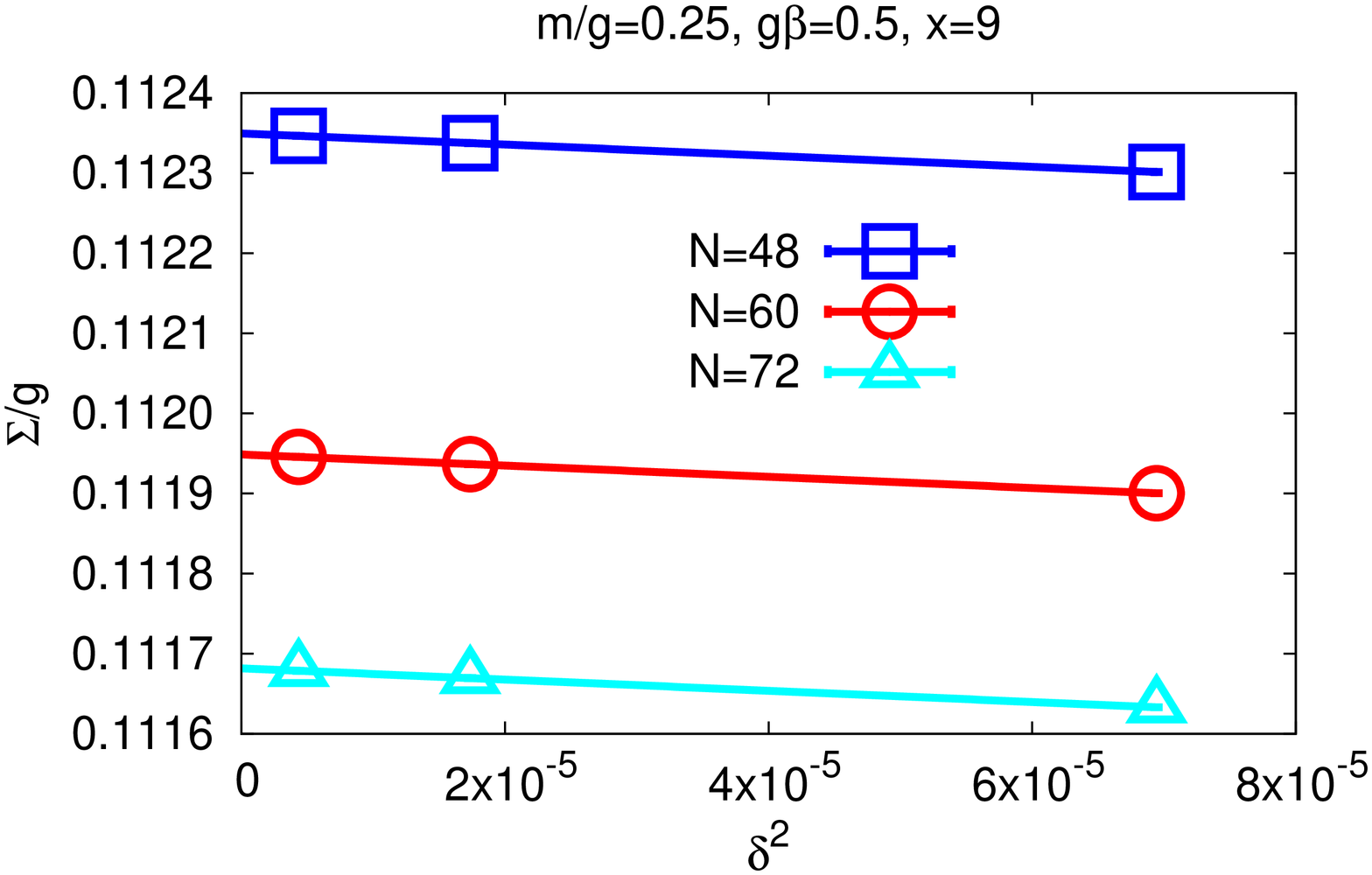}
\caption{\label{fig:D9}Examples of the $D$-dependence of the chiral condensate for $m/g=0.25$, $g\beta=0.5$, $L_{\rm cut}=10$, $x=9$, $N=60$ and $\delta=0.0020833$ (upper left), $\delta=0.0041667$ (upper right) and $\delta=0.0083333$ (lower left).
The red bands represent the uncertainty related to the bond dimension, taken as explained in the text.
Example of a $\delta$-extrapolation for $m/g=0.25$, $g\beta=0.5$, $L_{\rm cut}=10$, $x=9$ and three values of $N=48,\,60,\,72$ (lower right). Lines are fits of Eq.~(\ref{eq:fitdelta}). The data points have error bars, but they are too small to be seen.}
\end{center}
\end{figure}
\begin{figure}[t!]
\begin{center}
\includegraphics[width=0.345\textwidth,angle=270]{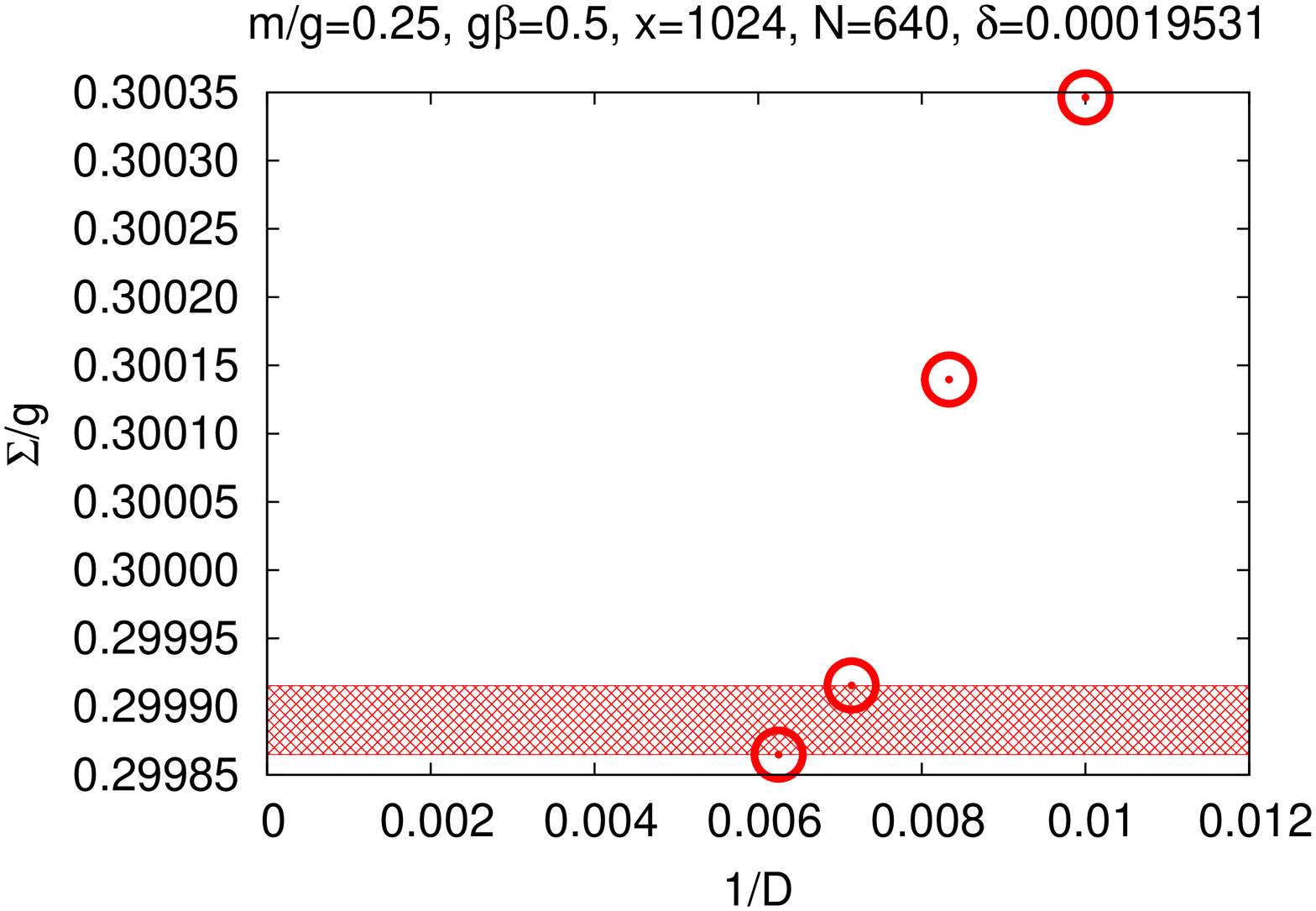}
\includegraphics[width=0.345\textwidth,angle=270]{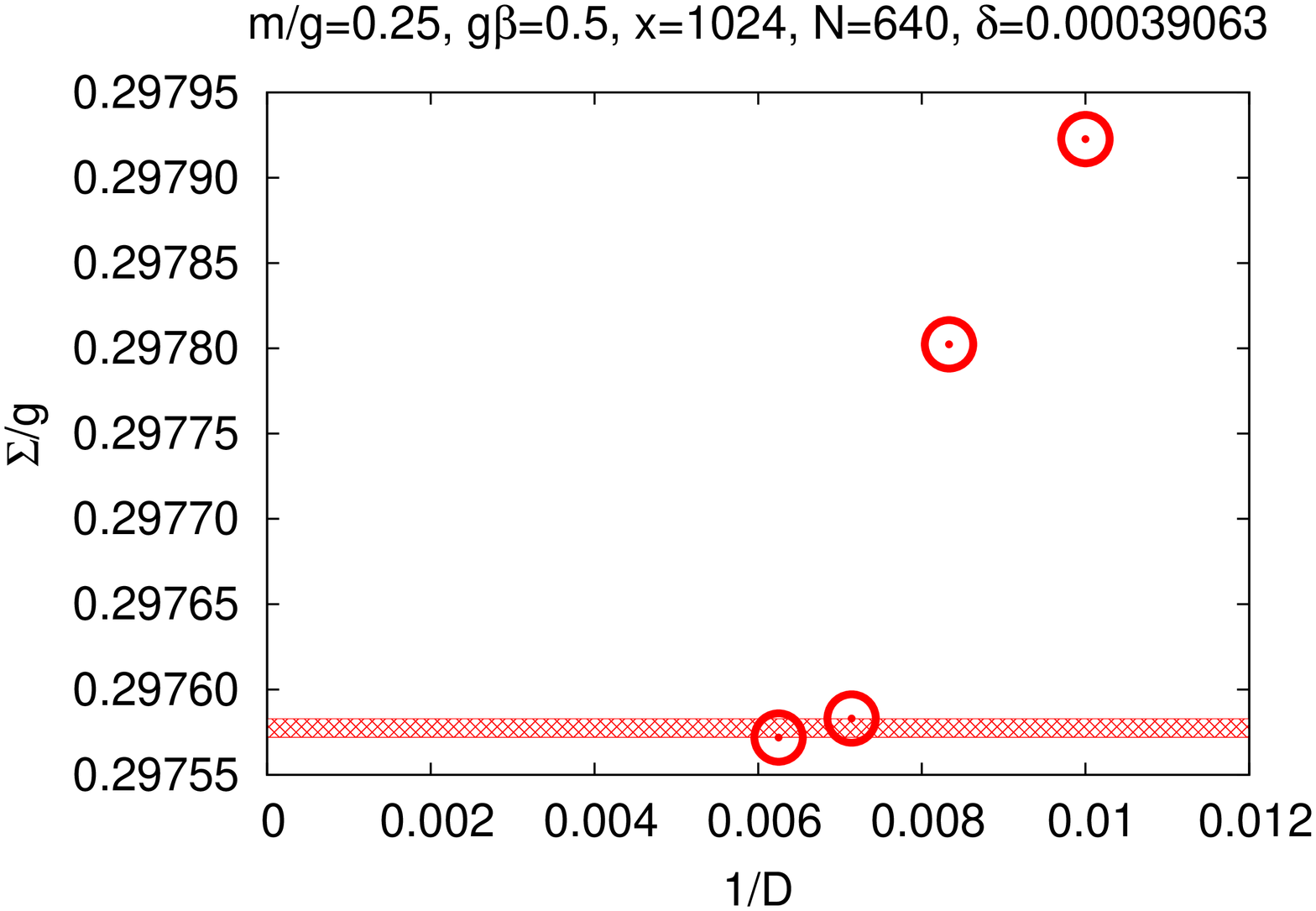}
\includegraphics[width=0.345\textwidth,angle=270]{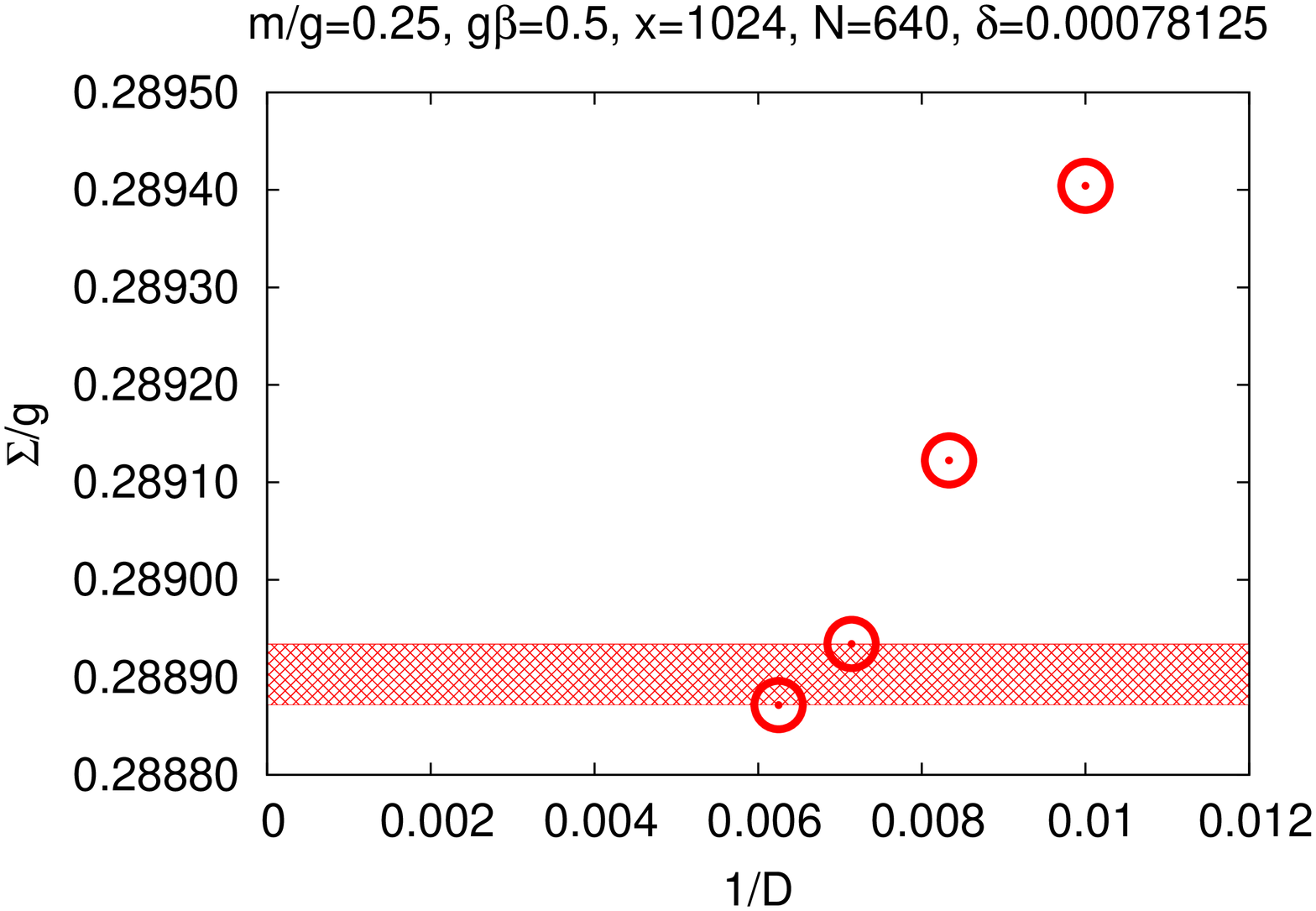}
\includegraphics[width=0.345\textwidth,angle=270]{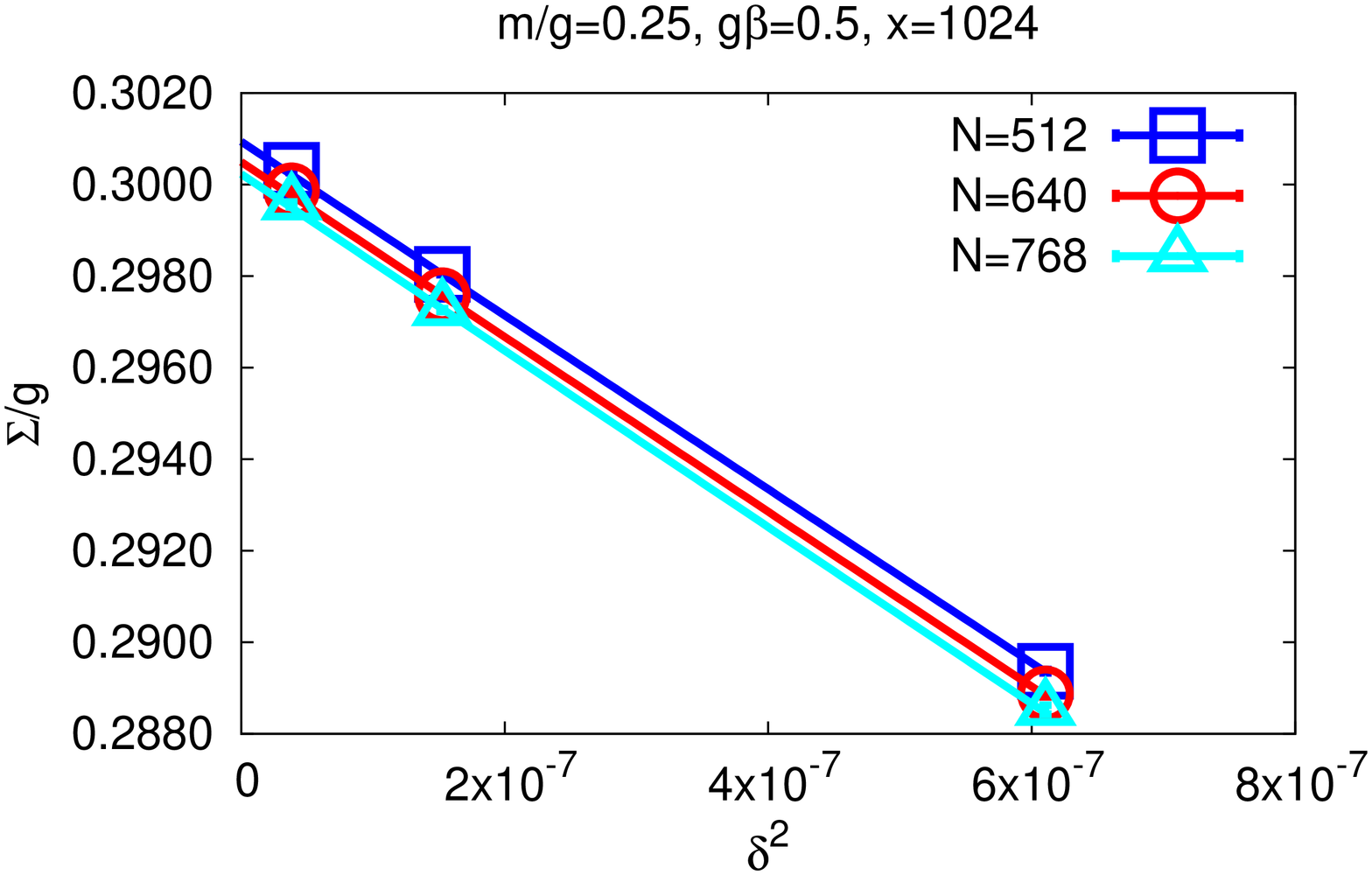}
\caption{\label{fig:D1024}Examples of the $D$-dependence of the chiral condensate for $m/g=0.25$, $g\beta=0.5$, $L_{\rm cut}=10$, $x=1024$, $N=640$ and $\delta=0.00019531$ (upper left), $\delta=0.00039063$ (upper right) and $\delta=0.00078125$ (lower left).
The red bands represent the uncertainty related to the bond dimension, taken as explained in the text.
Example of a $\delta$-extrapolation for $m/g=0.25$, $g\beta=0.5$, $L_{\rm cut}=10$, $x=1024$ and three values of $N=512,\,640,\,768$ (lower right). Lines are fits of Eq.~(\ref{eq:fitdelta}). The data points have error bars, but they are too small to be seen.}
\end{center}
\end{figure}

There are two new parameters with respect to $T=0$ computations, apart from the bond dimension, $D$, the system size, $N$, and the inverse coupling, $x$ --- the $L_{\rm cut}$ parameter describing the cut-off in the gauge sector and the step width, $\delta$. Thus, our sequence of extrapolations follows the order given below.

\textbf{Infinite bond dimension ($D\rightarrow\infty$) extrapolation.}
This extrapolation is done as in the $T=0$ case and we again take the result at our largest $D$ as the central value and the difference between this value and one at $(D-20)$ as the estimate of the uncertainty from the finite bond dimension.
Examples of such extrapolations are shown in Figs.~\ref{fig:D9} and \ref{fig:D1024}, for $x=9$ and $x=1024$, respectively (both at $m/g=0.25$, $g\beta=0.5$, $L_{\rm cut}=10$).
They illustrate a general feature in the $D$-dependence of the chiral condensate --- the convergence becomes worse towards the continuum limit.
However, this convergence is in all cases good --- the difference between our two largest bond dimensions (140 and 160) is of $\mathcal{O}(10^{-9})$ at $x=9$ and of $\mathcal{O}(5\cdot10^{-5})$ at $x=1024$.
This difference also depends on the temperature --- since lower temperatures are reached by increasing $g\beta$, the error from the finite bond dimension also increases at increasing $g\beta$, approximately linearly.
Note that in the thermal case, the convergence in $D$ is somewhat worse than at $T=0$ and we do not observe issues with insufficient machine precision (cf. Sec.~\ref{sec:zero} and the comments about double precision as not enough for certain parameter ranges).
Finally, there is little dependence on the value of $\delta$, the volume and on the fermion mass.
\\
\textbf{Zero step width ($\delta\rightarrow0$) extrapolation.}
We denote the results from the previous step as $\Sigma(m/g,x,N,\delta)$ and they differ from the $\delta=0$ limit by $\mathcal{O}(\delta^2)$.
Hence, we extrapolate to $\delta=0$ with:
\begin{equation}
\label{eq:fitdelta}
\Sigma(m/g,x,N,\delta)=\Sigma(m/g,x,N)+r(m/g,x,N)\delta^2,
\end{equation}
with the fitting parameters $\Sigma(m/g,x,N)$ and $r(m/g,x,N)$.
We always use three values of $\delta$ for each $(m/g,\,x,\,N)$, which allows us to verify that a fitting ansatz linear in $\delta^2$ is proper.
Since we want to access inverse temperatures $g\beta\in[0,8]$ with a step of $\Delta g\beta=0.1$, we use values of $\delta$ small enough such that this is possible.
Examples are shown in the lower right plots of Figs.~\ref{fig:D9} and \ref{fig:D1024}, for $x=9$ and $x=1024$, respectively (again at $m/g=0.25$, $g\beta=0.5$, $L_{\rm cut}=10$), and three volumes that are later used for infinite volume extrapolation.
Since the resulting errors are the propagated errors from the $D$-extrapolation, one again observes similar parameter dependences for the error obtained at this step. We also note that the linear ansatz (\ref{eq:fitdelta}) works very well.
\begin{figure}[t!]
\begin{center}
\includegraphics[width=0.345\textwidth,angle=270]{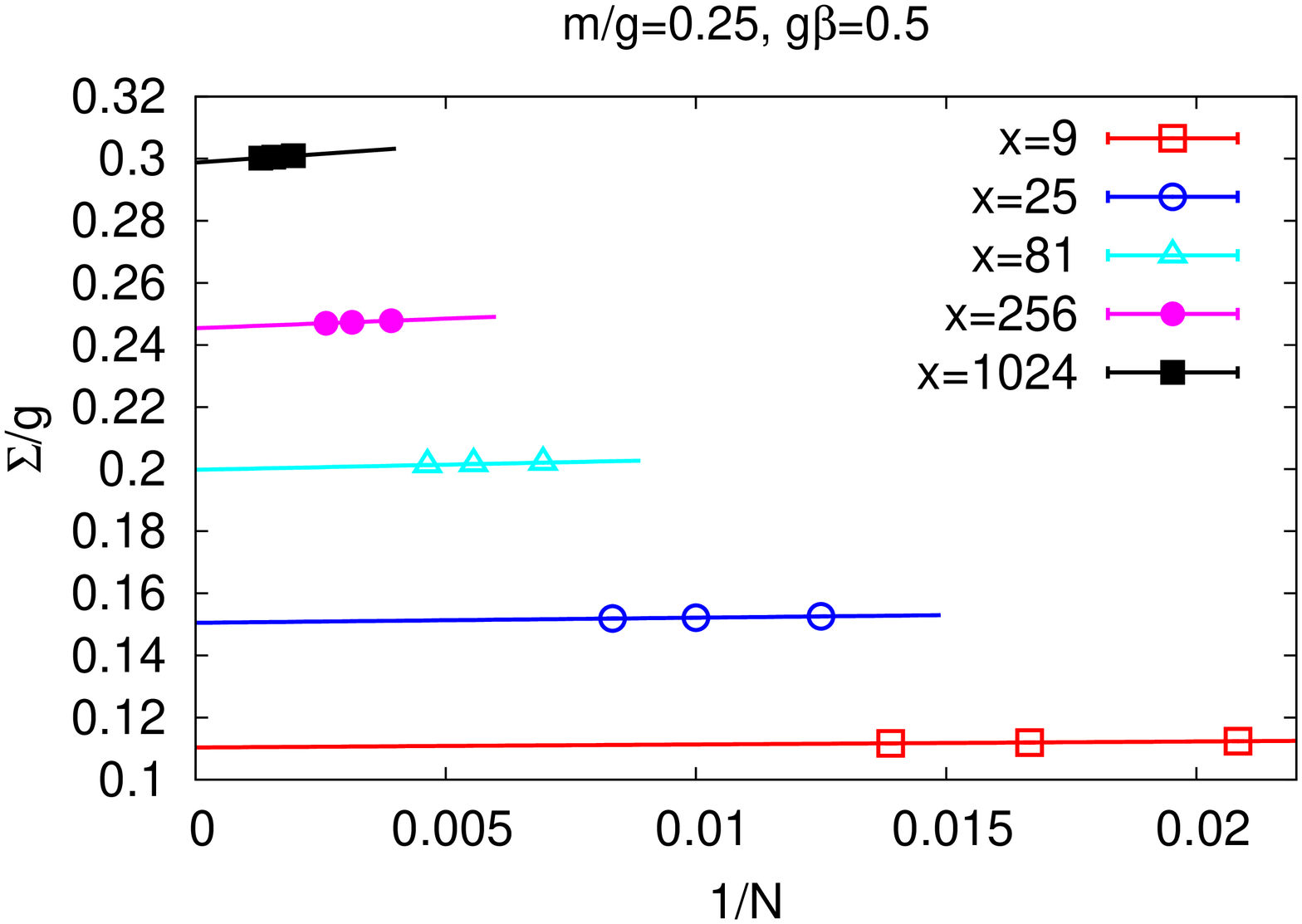}
\includegraphics[width=0.345\textwidth,angle=270]{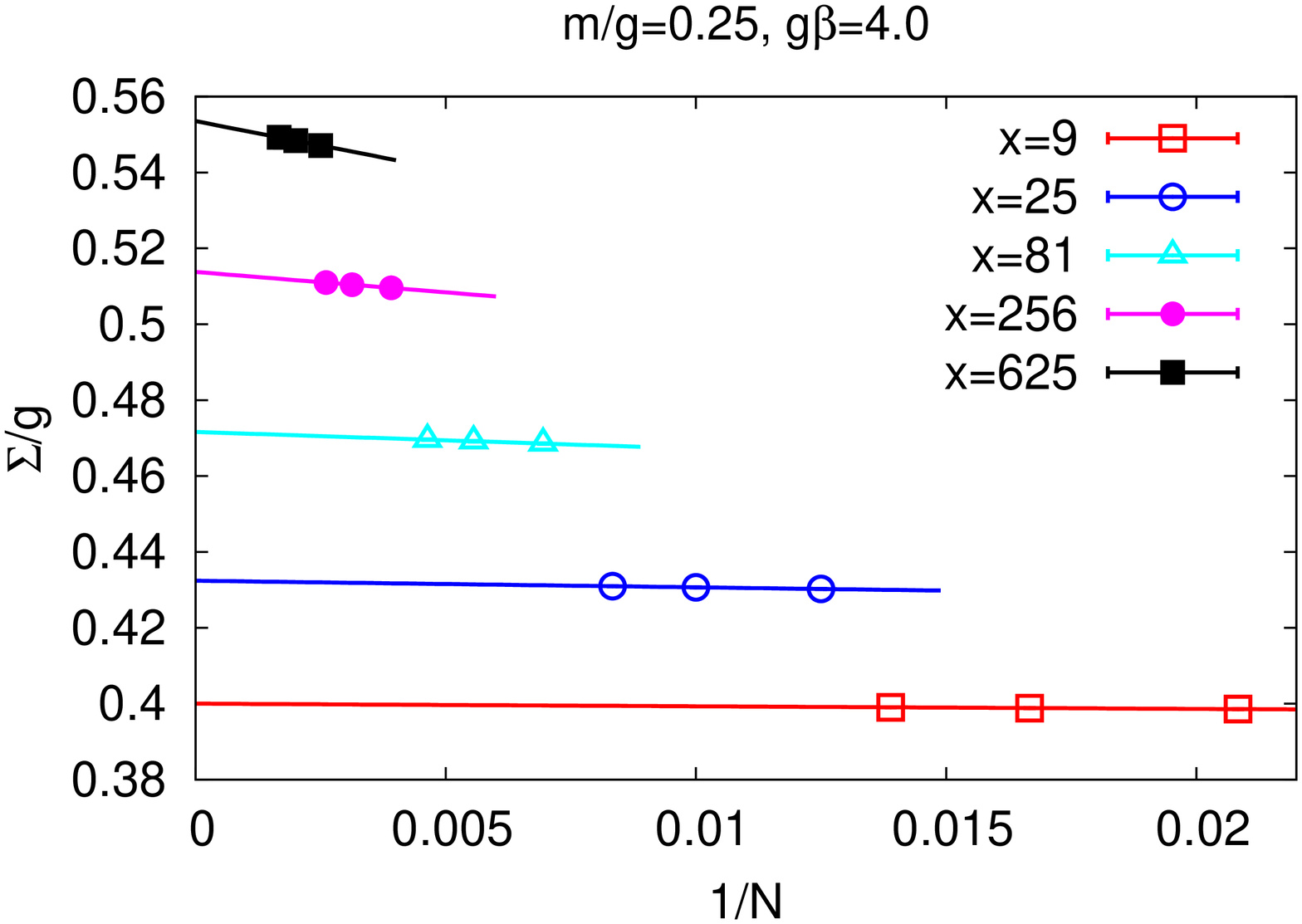}
\caption{\label{fig:N}Examples of the infinite volume extrapolations of the chiral condensate for $g\beta=0.5$ (left) and $g\beta=4$ and five values of $x$, with $L_{\rm cut}=10$. Lines are fits of Eq.~(\ref{eq:fitN}).
}
\end{center}
\end{figure}
\\
\textbf{Infinite volume ($N\rightarrow\infty$) extrapolation.}
The results corresponding to our estimates of the $D\rightarrow\infty$ and $\delta\rightarrow0$ limits are extrapolated to infinite volume by using the same kind of linear fitting ansatz as in the $T=0$ case, i.e. Eq.~(\ref{eq:fitN}), and volumes $N=\{16\sqrt{x},\,20\sqrt{x},\,24\sqrt{x}\}$. 
An example extrapolation is shown in Fig.~\ref{fig:N}, for $m/g=0.25$, $L_{\rm cut}=10$, five values of the lattice spacing and two temperatures: $g\beta=0.5$ (left) and $g\beta=4$ (right).
As in the $T=0$ case, we observe that the fitting ansatz gives very good description of our data.
\\
\textbf{Removing the $L_{\rm cut}$ cut-off ($L_{\rm cut}\rightarrow\infty$ extrapolation).}
The physical results have to be independent of the used gauge sector cut-off.
We found empirically that for all ranges of our parameters, $L_{\rm cut}=10$ always yields results compatible with $L_{\rm cut}=8$ and $L_{\rm cut}=12$.
Hence, this value of $L_{\rm cut}$ is effectively $L_{\rm cut}=\infty$ and no explicit extrapolation is needed (see also Ref.~\cite{Banuls:2015sta}).
\\
\textbf{Continuum limit ($x\rightarrow\infty$) extrapolation.}
As our final step, we perform the continuum limit extrapolation of the infinite volume results $\Sigma(m/g,x)$.
Before this is done, we subtract the infinite volume free condensate according to Eq.~(\ref{eq:subtr}) and obtain the subtracted condensate $\Sigma_{\rm subtr}(m/g,x)$.
We consider the following three fitting ansatzes:
\begin{equation}
\label{eq:cont1} 
\Sigma_{\rm subtr}(m/g,x)=\Sigma_{\rm subtr}^{(1)}(m/g)+\frac{a^{(1)}(m/g)}{\sqrt{x}}\log(x)+\frac{b^{(1)}(m/g)}{\sqrt{x}},
\end{equation}
\begin{equation}
\label{eq:cont2} 
\Sigma_{\rm subtr}(m/g,x)=\Sigma_{\rm subtr}^{(2)}(m/g)+\frac{a^{(2)}(m/g)}{\sqrt{x}}\log(x)+\frac{b^{(2)}(m/g)}{\sqrt{x}}+\frac{c^{(2)}(m/g)}{x},
\end{equation}
\begin{equation}
\label{eq:cont3} 
\Sigma_{\rm subtr}(m/g,x)=\Sigma_{\rm subtr}^{(3)}(m/g)+\frac{a^{(3)}(m/g)}{\sqrt{x}}\log(x)+\frac{b^{(3)}(m/g)}{\sqrt{x}}+\frac{c^{(3)}(m/g)}{x}+\frac{d^{(3)}(m/g)}{x^{3/2}},
\end{equation}
which differ by the order of the polynomial in $1/\sqrt{x}$. We refer to them as linear+log, quadratic+log and cubic+log, respectively.
We observe that the discretization effects are very different at different temperatures, in particular these effects become very strong at high temperatures and a polynomial cubic in $1/\sqrt{x}$ is needed to obtain a good description of data.
We adopt a modified procedure to obtain the systematic error from the choice of the fitting range and the fitting ansatz.
The procedure used to analyze the $T=0$ data is inappropriate here, because of the large dependence of the uncertainty from the $D$-extrapolation on the lattice spacing. This uncertainty at a fine lattice spacing ($x=100-500$) is up to four orders of magnitude larger than the one for our coarsest lattice spacings.
Hence, the analogue of the weighted histogram built at $T=0$ is no longer reliable, as it contains fits with very large uncertainties.
This does not happen at $T=0$, where the fine lattice spacings have only slightly larger uncertainties from the $D$ and $N$-extrapolations than the coarse lattice spacings.
This reflects the difference in strategies used to approximate thermal and ground states as tensor networks.
In practice, it translates into a somewhat different manner the truncation errors are accumulated in the thermal evolution with respect to the $T=0$ algorithm.
At large $g\beta$, i.e. after several steps of imaginary time evolution, the truncation errors are much larger than in the ground state.
As a consequence, the $T=0$ procedure of obtaining the systematic error does not make sense in the $T>0$ case, since only one or two fits dominate the weighted histogram.

For this reason, the procedure to extract the fitting range/ansatz uncertainty is the following.
It is performed separately for each temperature $g\beta$ at a given fermion mass $m/g$.
We fix the maximum $x$ entering each fit to be the one corresponding to the finest lattice spacing.
Then, we build all possible fits of Eqs.~(\ref{eq:cont1})-(\ref{eq:cont3}) changing only the minimal entering $x$ ($x_{\rm min}$).
We take as the central value $\Sigma_{\rm subtr}^{(i)}(m/g)$ that corresponds to the smallest uncertainty propagated through $D$, $\delta$ and $N$-extrapolations, but one that satisfies the condition $\chi^2/{\rm dof}\leq1$ and has all its fitting coefficients statistically significant.
We denote it by $\Sigma_{\rm subtr}(m/g)$ and its error by $\Delta\Sigma_{\rm subtr}(m/g)$.
We combine this uncertainty quadratically with the uncertainty from the choice of the fitting interval, $\Delta^{\rm interval}\Sigma_{\rm subtr}(m/g)$, and from the choice of the fitting ansatz, $\Delta^{\rm ansatz}\Sigma_{\rm subtr}(m/g)$.
The former is defined as the difference between $\Sigma_{\rm subtr}(m/g)$ and the most outlying $\Sigma_{\rm subtr}^{(i)}(m/g)$ (corresponding to the same $(i)$, i.e. the same functional form of the fitting ansatz) which has still all the fitting coefficients statistically significant.
The latter is taken to be the difference between $\Sigma_{\rm subtr}(m/g)$ and the most outlying $\Sigma_{\rm subtr}^{(j)}(m/g)$ (where $(j)\neq(i)$, i.e. from another fitting ansatz) which has again statistically significant fitting coefficients.

\begin{figure}[t!]
\begin{center}
\includegraphics[width=0.345\textwidth,angle=270]{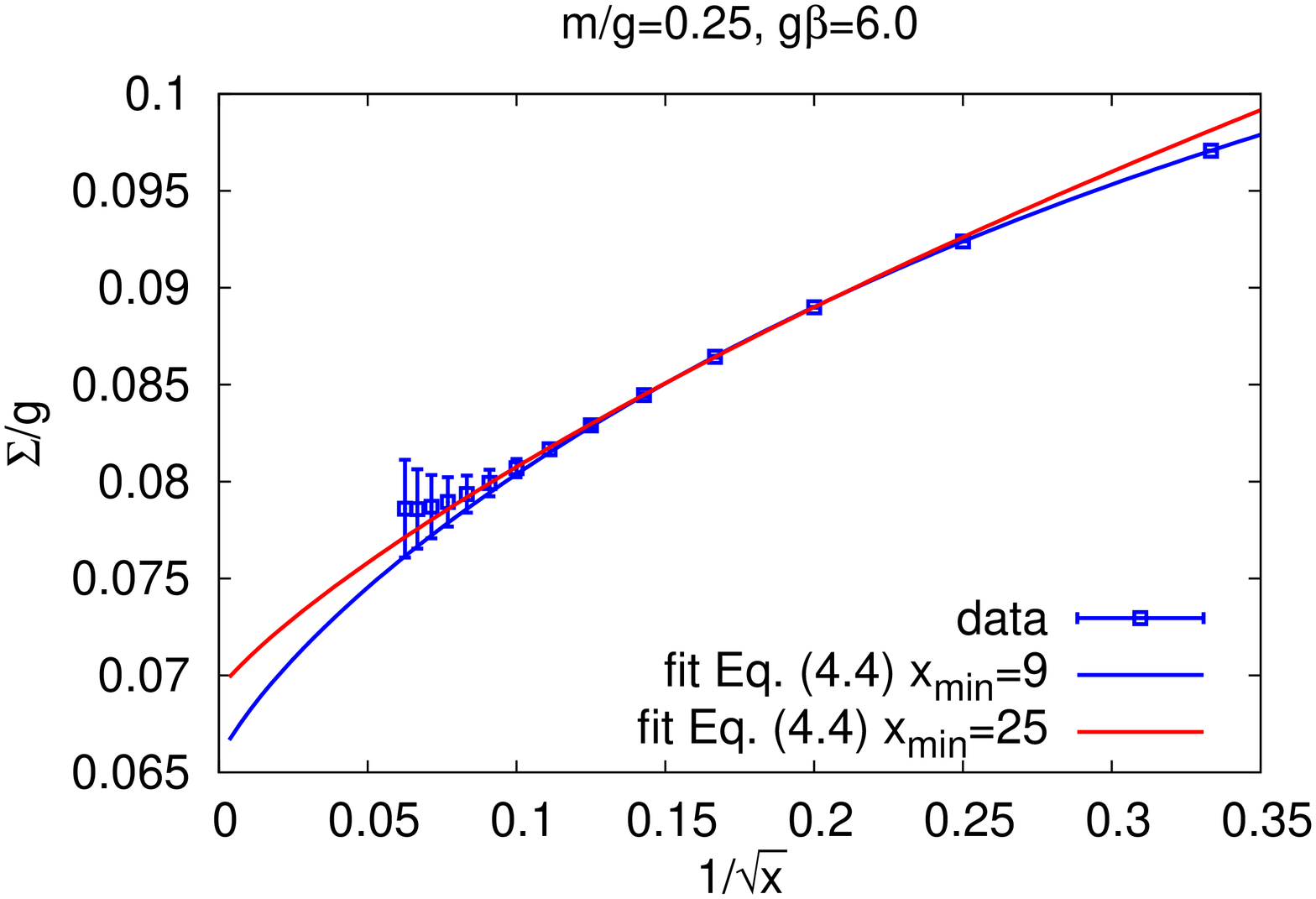}
\includegraphics[width=0.345\textwidth,angle=270]{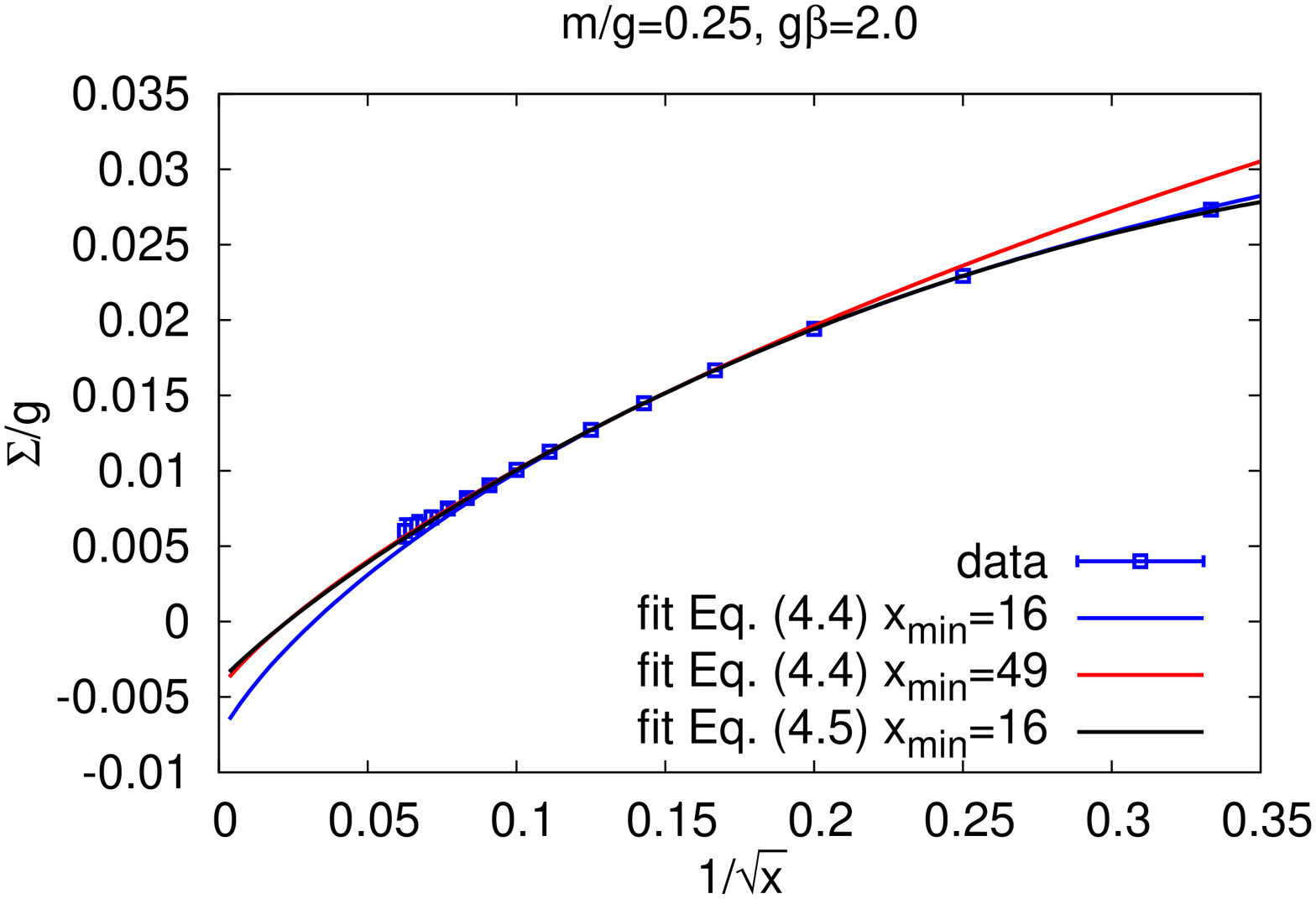}
\includegraphics[width=0.345\textwidth,angle=270]{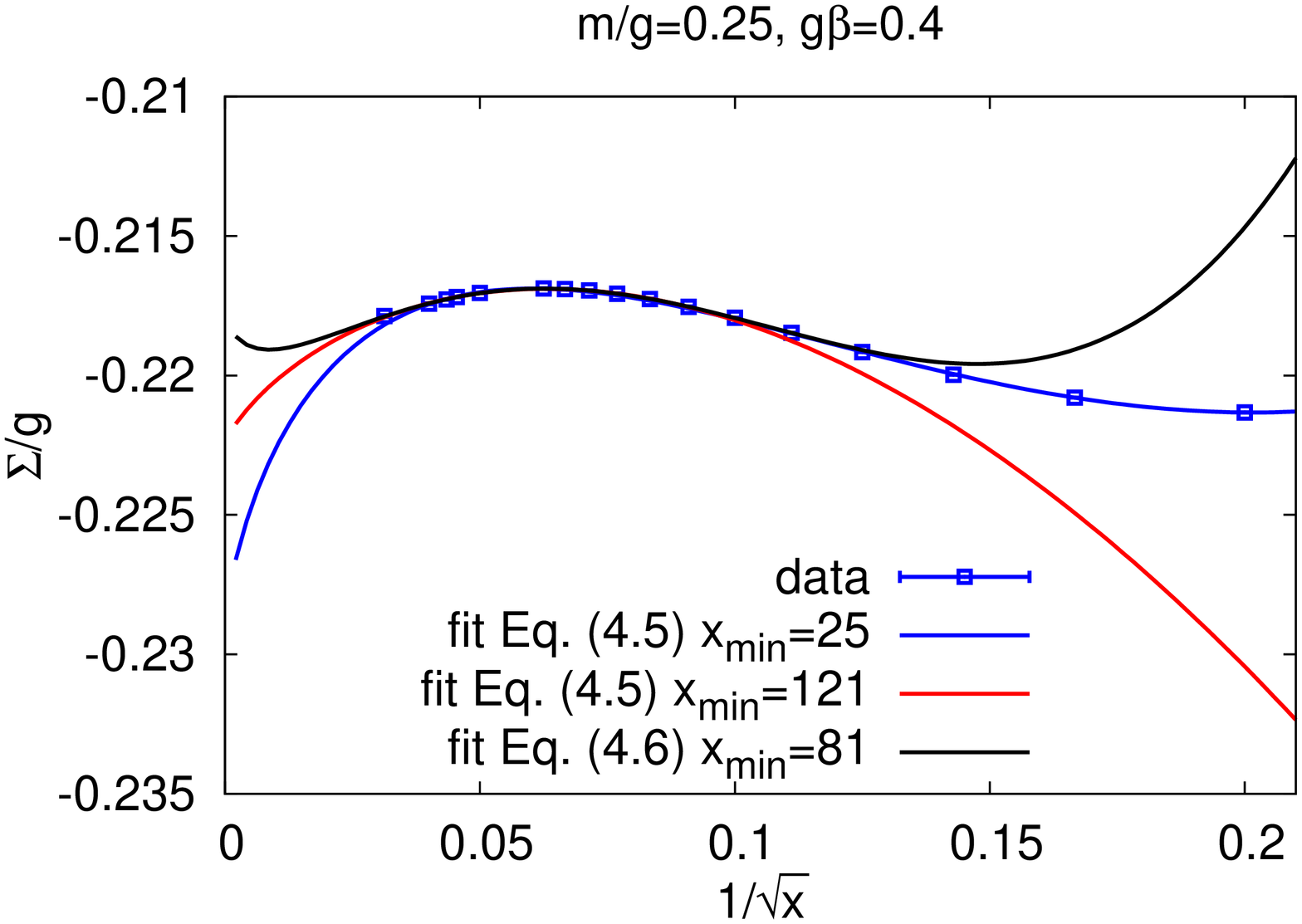}
\includegraphics[width=0.345\textwidth,angle=270]{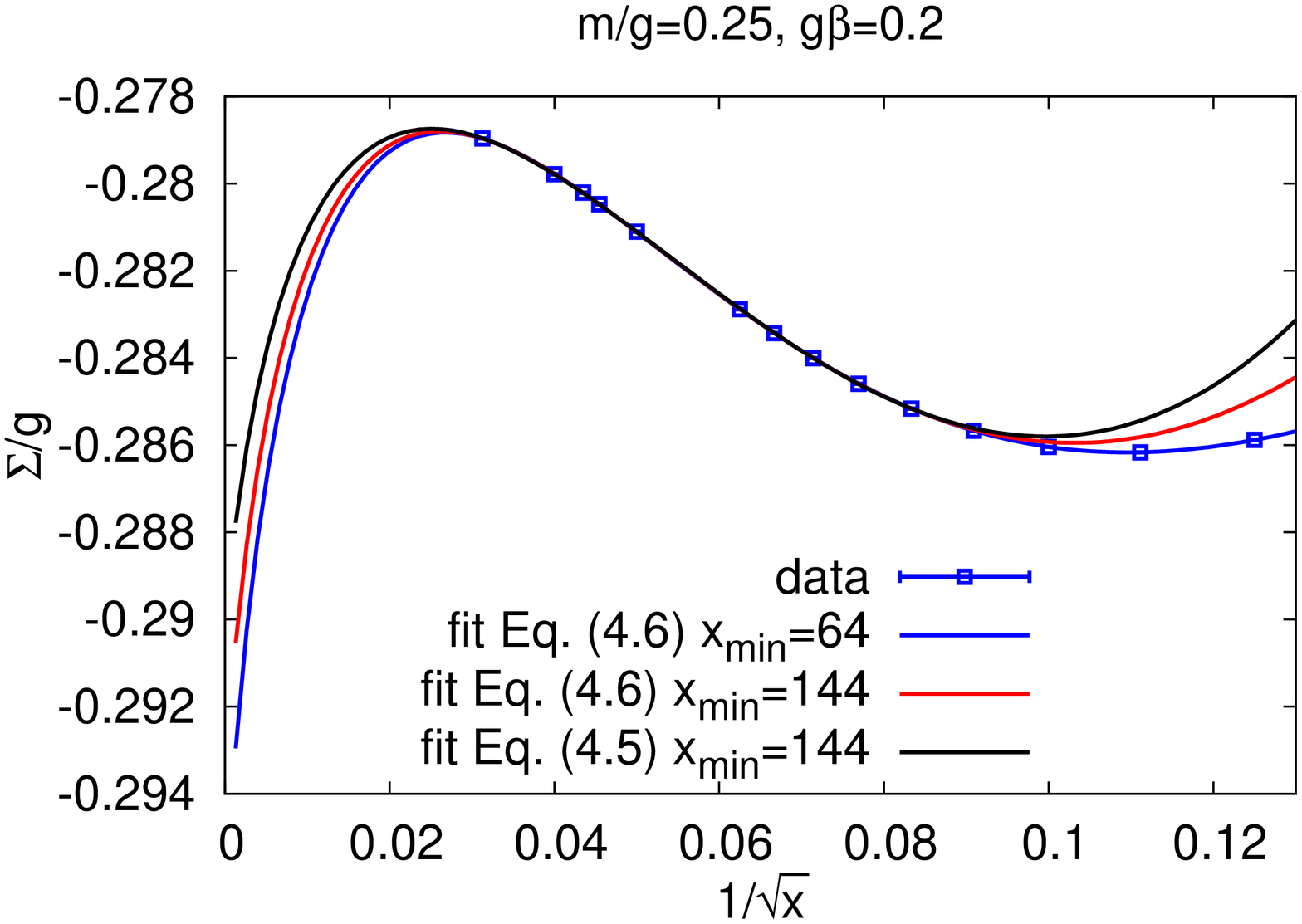}
\caption{\label{fig:cont}Examples of the continuum limit extrapolations of the chiral condensate for $m/g=0.25$ and temperatures $g\beta=6.0$ (upper left), $g\beta=2.0$ (upper right), $g\beta=0.4$ (lower left) and $g\beta=0.2$ (lower right).
}
\end{center}
\end{figure}

Below, we illustrate this procedure with a few examples at the fermion mass $m/g=0.25$ (Fig.~\ref{fig:cont}).
We start with a low temperature, $g\beta=6$, effectively corresponding to $T=0$ (after a certain $m$-dependent $g\beta$, the continuum result does not change any more --- in the case of $m/g=0.25$, zero temperature is reached around $g\beta=6$).
Here, taking the linear+log fitting ansatz and $x_{\rm min}=9$ yields a good fit, with $\chi^2/{\rm dof}\approx0.07$. It can be compared to only two other fits, both of them linear+log, with $x_{\rm min}=16$ and $x_{\rm min}=25$. Increasing $x_{\rm min}$ further or changing the fit form to quadratic+log or cubic+log leads to at least one of the fitting coefficients becoming statistically insignificant.
Hence, our final result for this temperature and fermion mass is $\Sigma_{\rm subtr}(\Delta)(\Delta^{\rm interval})(\Delta^{\rm ansatz})=0.0657(3)(43)(0)$ and is dominated by the uncertainty from the choice of the fitting interval. The error from the choice of the fitting ansatz is zero, since no quadratic+log or cubic+log fit produces a significant result. 
Since $g\beta=6$ is effectively $T=0$, this result can be compared to our $T=0$ result at this fermion mass in Tab.~\ref{tab:T0}.
We observe full consistency, although the precision of the thermal computation is four orders of magnitude worse than of the ground state one.
This is hardly surprising, as thermal evolution is definitely not the best method to investigate ground state properties.

Another example continuum extrapolation is shown for $g\beta=2$ (upper right plot of Fig.~\ref{fig:cont}). In this case, the central value comes from a linear+log fit with $x_{\rm min}=16$ and it is compared to the same functional form of the fit with $x_{\rm min}=49$ as well as to a quadratic+log fit with $x_{\rm min}=16$. Finally, we get $\Sigma_{\rm subtr}(\Delta)(\Delta^{\rm interval})(\Delta^{\rm ansatz})=-0.0078(0.3)(36)(38)$. 
Towards higher temperatures, cut-off effects become increasingly important, in the sense that one needs higher order polynomials in $1/\sqrt{x}$.
For $g\beta=0.4$ (lower left of Fig.~\ref{fig:cont}), the central value that we take comes from a quadratic+log fit with $x_{\rm min}=25$, compared to $x_{\rm min}=121$ and a cubic+log fit with $x_{\rm min}=81$. This leads to $\Sigma_{\rm subtr}(\Delta)(\Delta^{\rm interval})(\Delta^{\rm ansatz})=-0.229(0.05)(6)(11)$.
Our final example is $g\beta=0.2$ (lower right of Fig.~\ref{fig:cont}). Here, the central value comes from a cubic+log fit with $x_{\rm min}=64$, compared to $x_{\rm min}=144$ and a quadratic+log fit with $x_{\rm min}=144$. We get $\Sigma_{\rm subtr}(\Delta)(\Delta^{\rm interval})(\Delta^{\rm ansatz})=-0.297(0.14)(3)(7)$.
In all these cases, the error is dominated by the uncertainty from the choice of the fitting interval and ansatz.
Nevertheless, with the adopted systematic error estimation procedure, one can have these uncertainties reliably under control.

\begin{figure}[t!]
\begin{center}
\includegraphics[width=0.345\textwidth,angle=270]{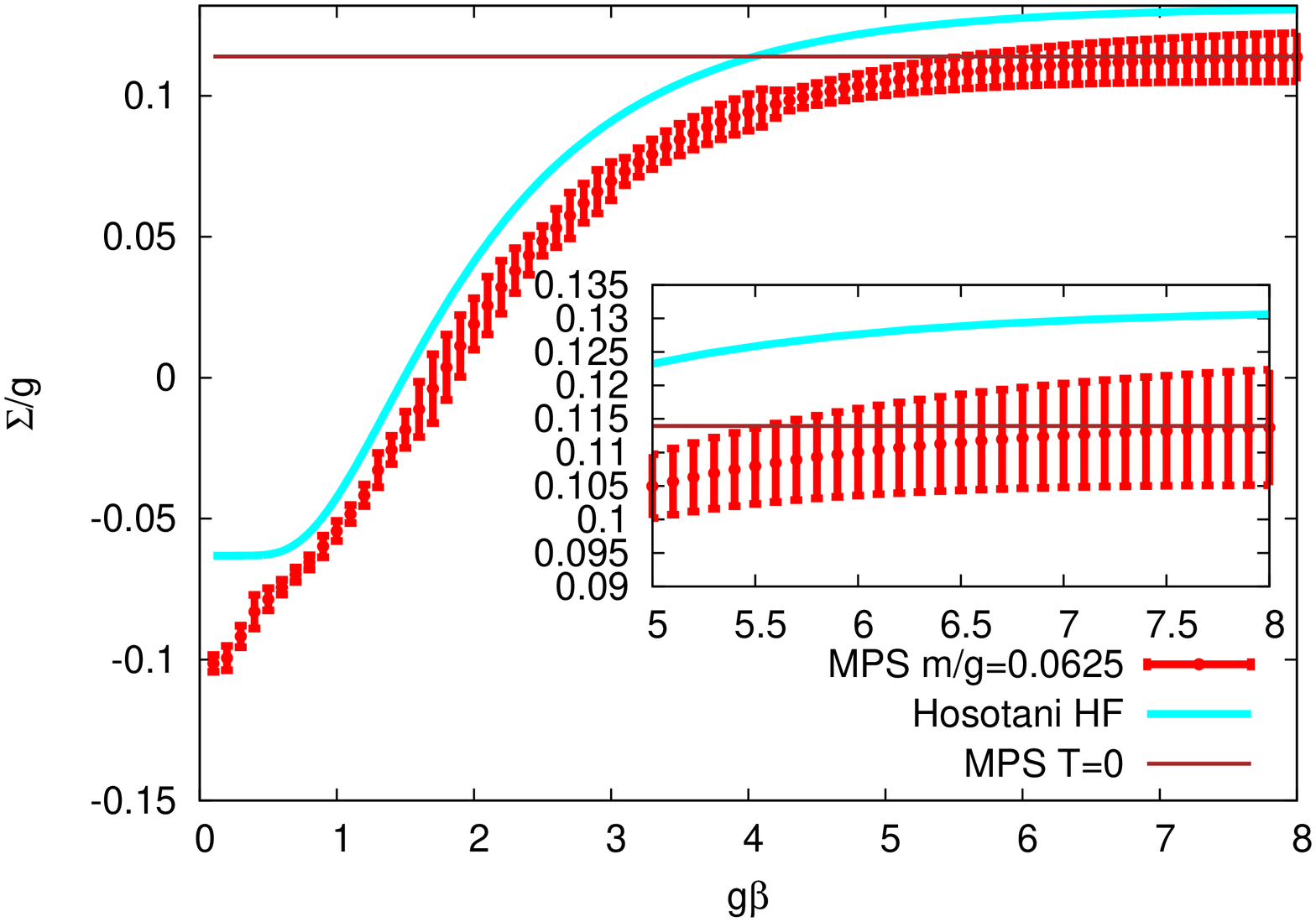}
\includegraphics[width=0.345\textwidth,angle=270]{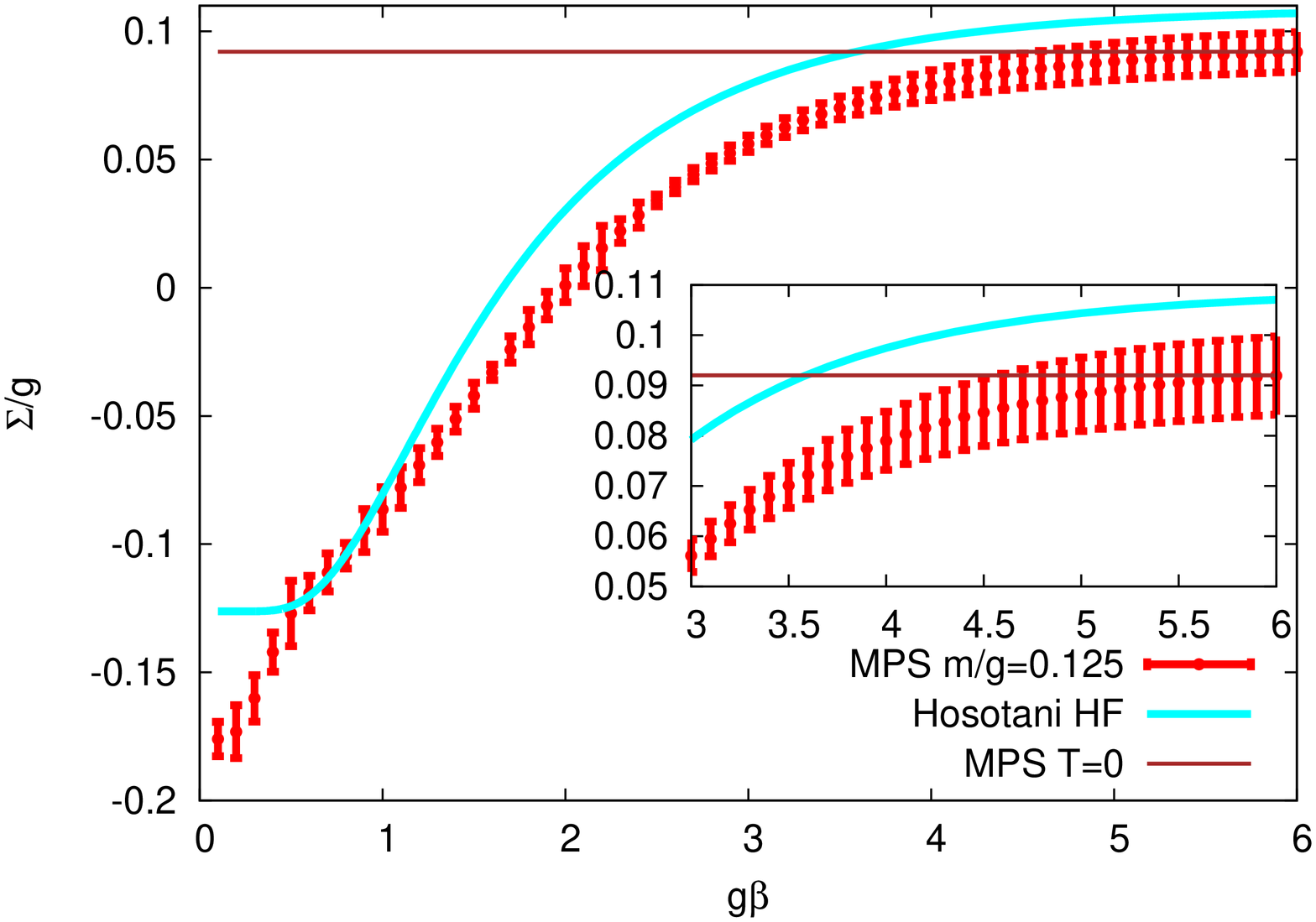}
\includegraphics[width=0.345\textwidth,angle=270]{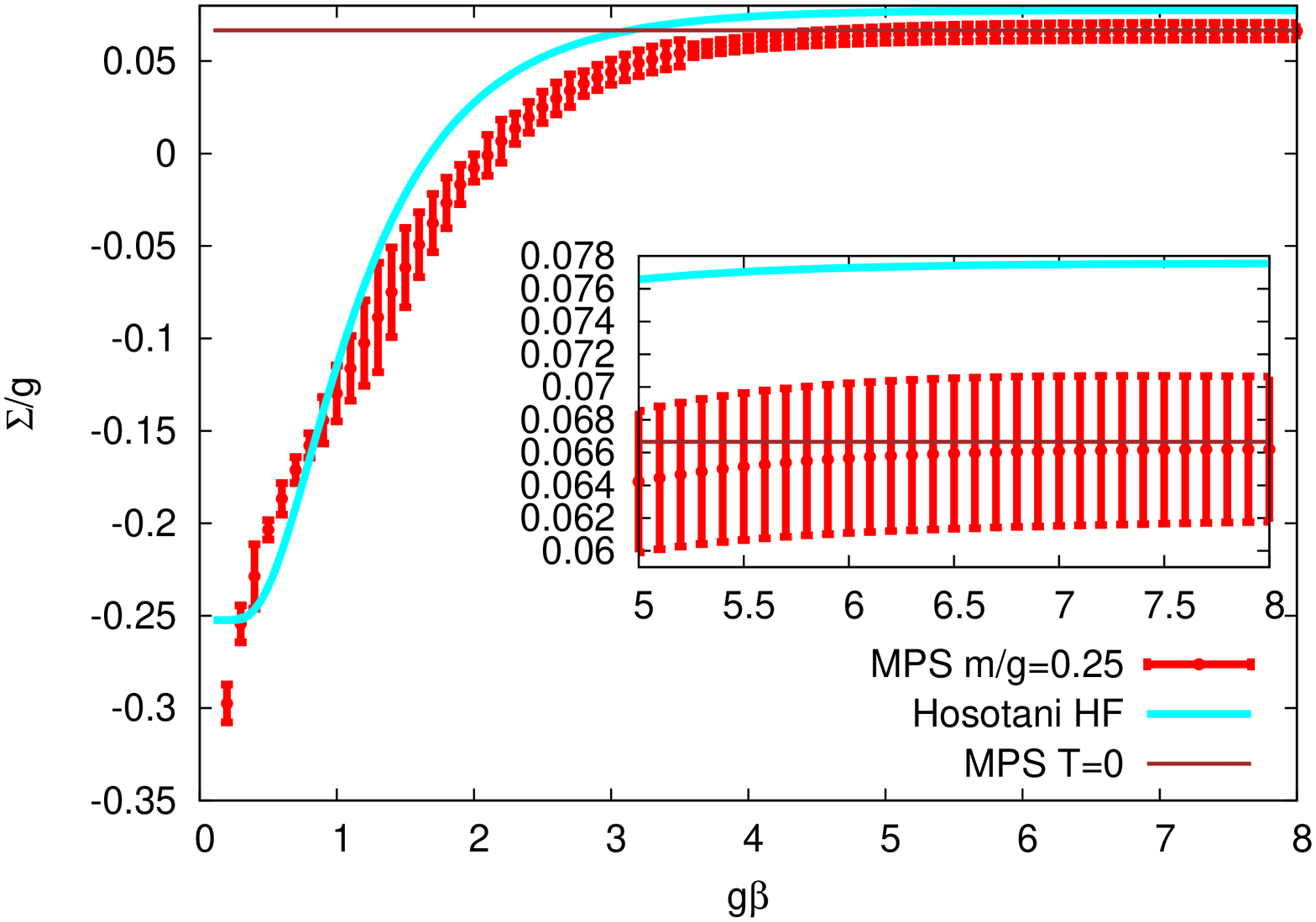}
\includegraphics[width=0.345\textwidth,angle=270]{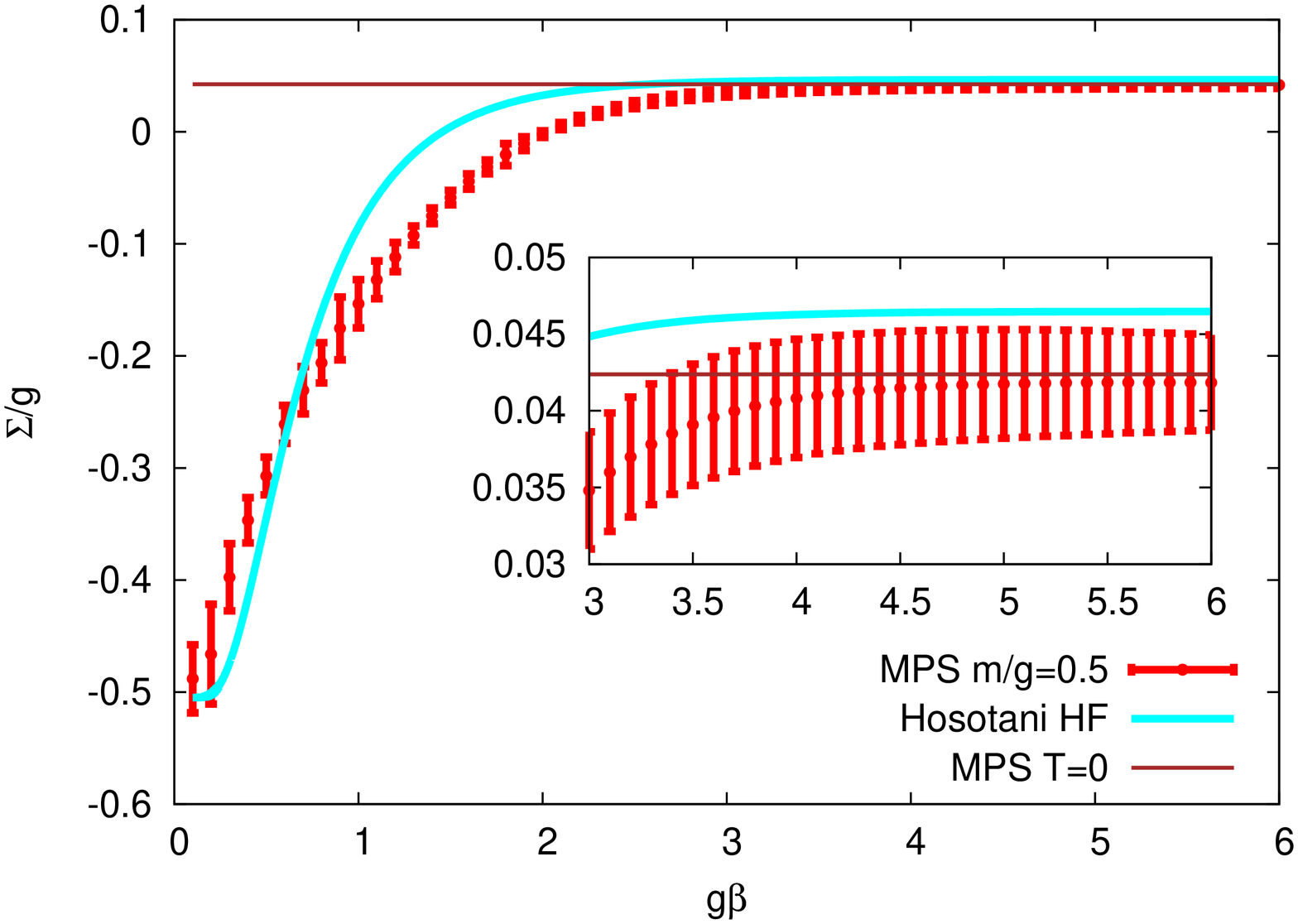}
\includegraphics[width=0.345\textwidth,angle=270]{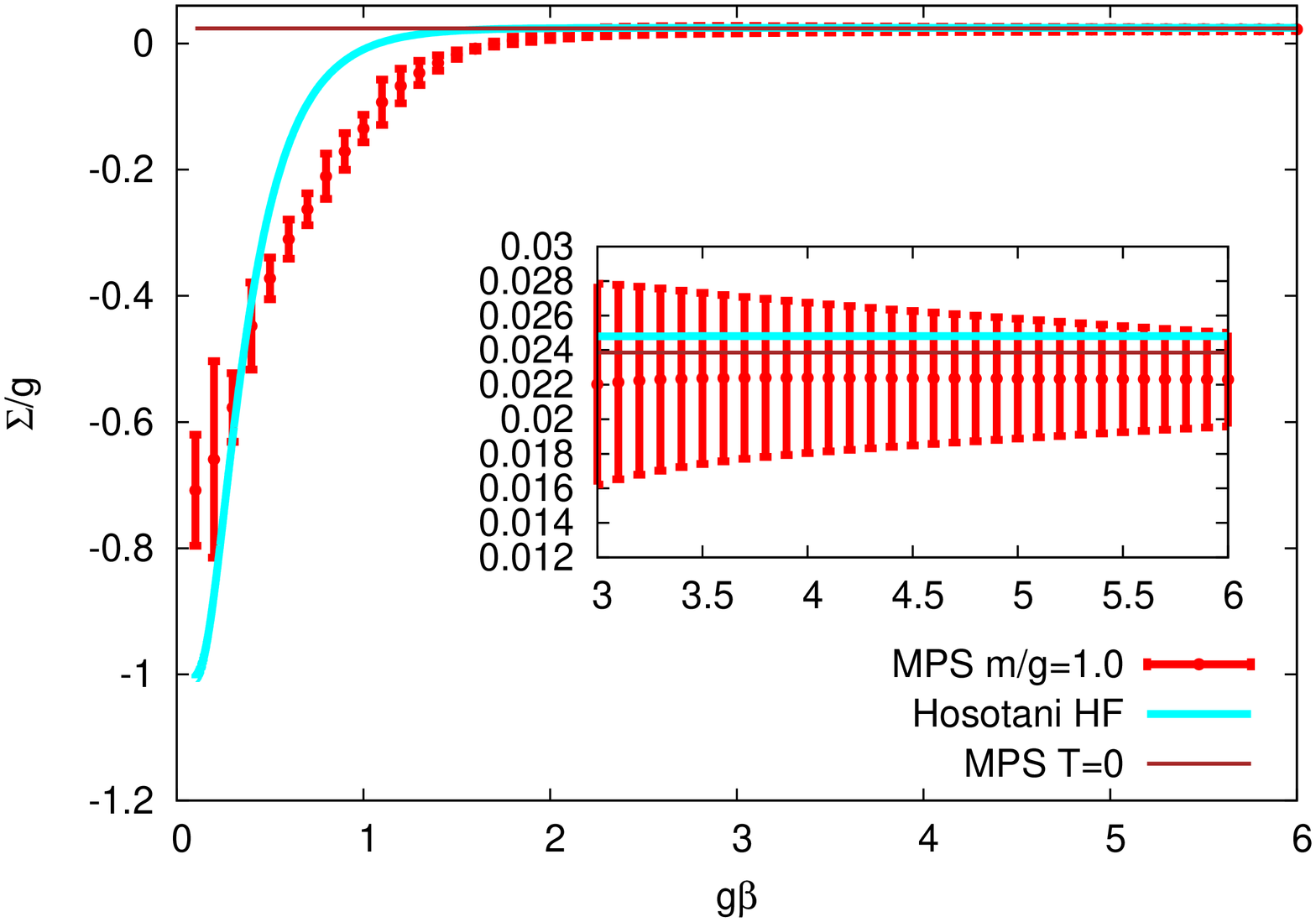}
\caption{\label{fig:gbeta}Inverse temperature dependence of the continuum limit extrapolated chiral condensate for all our fermion masses. Shown is also the result obtained at $T=0$ and the result of the approximation of Ref.~\cite{Hosotani:1998za}.}
\end{center}
\end{figure}

We repeat the analysis steps for all our fermion masses and we summarize the continuum limit results in Fig.~\ref{fig:gbeta}, where we show results up to $g\beta=8$ ($m/g=0.0625$ and $0.25$) or $g\beta=6$ ($m/g=0.125,\,0.5,\,1$).
The most important feature confirming the validity of our results is that we always reproduce the $T=0$ result within our errors --- actually the difference between our central values at large enough $g\beta$ and the $T=0$ MPS result is much smaller than our errors, suggesting that the error estimation procedure is rather conservative. 
We also note that our systematic error procedure makes the final errors strongly dependent on temperature --- with sometimes irregular jumps of the error caused by some other fitting interval or fitting ansatz entering the procedure at certain $g\beta$ values\footnote{For example, at $m/g=0.25$ all the quadratic+log fits have at least one fitting coefficient statistically insignificant above $g\beta=3.5$ and at this temperature and higher (smaller $g\beta$), quadratic+log fits become statistically significant and thus enlarge our error.}.
Apart from the agreement with the $T=0$ result, we observe that the approach to this result is faster for higher fermion masses --- for $m/g=1$, $g\beta=3$ is already effectively zero temperature, while for our lowest mass, $m/g=0.0625$, we have small changes of the central value even above $g\beta=6$.
Concerning the agreement with the approximation of Ref.~\cite{Hosotani:1998za} (referred to as ``Hosotani HF'' in the plot), the latter provides good qualitative description of the temperature dependence of the chiral condensate.
However, the quantitative agreement is not perfect, with typical deviations of 10-20\%.
It is known that the approximation becomes exact in the massless limit and indeed, e.g. Hosotani's $T=0$ result at $m/g=0.0625$ is relatively closer to the MPS result than the one at $m/g=0.125$.
On the other hand, the approximation of Ref.~\cite{Hosotani:1998za} also approaches the analytical result of zero at infinite fermion mass and $T=0$ --- hence one also expects an increasing agreement in this regime.
Indeed, the relative difference at $m/g=1$ is the smallest from among all our considered masses.
However, when we consider the slope of the $g\beta$-dependence, we clearly observe that the agreement between Hosotani HF and our computation becomes better towards small fermion masses, with both curves being almost parallel for $m/g=0.0625$.

\section{Summary and prospects}
\label{sec:summary}

In this paper, we have performed a study of the temperature dependence of the 
chiral condensate for the one-flavour Schwinger model using a Hamiltonian approach.   
We emphasize that while for zero temperature we employ 
a matrix product {\em state} (MPS) ansatz, for 
non-vanishing temperature we use a 
matrix product {\em operator} (MPO) ansatz.    
In addition, for the non-zero temperature calculation, we have to perform
a thermal evolution by starting from a well defined infinite temperature 
state and evolve the system in incremental inverse temperature steps 
towards zero temperature using a density operator.        

Thus, non-zero temperature calculations within the Hamiltonian approach 
are rather different from the so far carried out zero temperature ones and 
hence non-zero temperature computations for gauge 
theories are novel and need to be tested. 
While in Ref.~\cite{Banuls:2015sta} we have initiated such non-zero temperature 
computations for massless fermions, in this paper we went substantially beyond 
this work by studying the system at various fermion masses. In addition, we employed 
consistently a truncation of the gauge sector. This allowed us to 
reach very large system sizes and, keeping the physical extent of the 
model fixed, very small values of the lattice spacing. 

Within our calculation of the chiral condensate, we carried out a substantial
and challenging effort to control the systematic effects. To this end, we 
performed extrapolations to zero thermal evolution step size, infinite 
bond dimension, infinite volume and zero lattice spacing. In addition, we
tested that our cut parameter for the gauge sector truncation has been 
sufficiently large. The 
final non-trivial check of the validity of our approach has been 
to recover the zero temperature result of the chiral condensate after 
the long thermal evolution performed. 

As a result of our work, we could compute the chiral condensate over a broad temperature range 
from infinite to almost zero temperature with controlled errors. 
This has been done for zero, light and heavy fermion masses. 
For zero fermion mass, we found excellent agreement with the analytical 
results of Ref.~\cite{Sachs:1991en}. Moving to non-zero fermion masses, 
a comparison to Ref.~\cite{Hosotani:1998za} did not lead to a clear 
conclusion, see Fig.~\ref{fig:gbeta}. 
Although qualitatively the temperature dependence of the chiral condensate
shows a comparable behaviour between the analytical result of Ref.~\cite{Hosotani:1998za}
and our data, there does not seem to be an agreement on the quantitative level. 
This is presumably due to the fact that the approximations made in Ref.~\cite{Hosotani:1998za} are too rough to reach a satisfactory quantitative 
agreement. 

We consider the here performed work, besides of the clear interest in its own, as a 
necessary step towards investigating the Schwinger model 
when adding a chemical potential. This setup 
leads to the infamous sign problem and it would be very reassuring to see whether 
the here used MPS and MPO approaches can lead to a successful application 
for this very hard problem, which is very difficult, if not impossible to solve by standard 
Markov chain Monte Carlo methods.

\begin{acknowledgments}
We thank J. I. Cirac for discussions.
This work was partially funded by the EU through SIQS grant (FP7 600645).
K.C. was supported in part by the Helmholtz International Center for FAIR within the framework of the LOEWE program launched by the State of Hesse and in part by the Deutsche Forschungsgemeinschaft (DFG), project nr. CI 236/1-1 (Sachbeihilfe).
Calculations for this work were performed on the LOEWE-CSC high-performance computer of Johann Wolfgang Goethe-University Frankfurt am Main and in the computing centers of DESY Zeuthen and RZG Garching.
\end{acknowledgments}

\bibliography{MPSSchwinger}
\bibliographystyle{JHEP}

\end{document}